\documentstyle[12pt]{article}

\begin{document}

\renewcommand{\theequation}{\thesection.\arabic{equation}}
\renewcommand{\thefootnote}{\fnsymbol{footnote}}

\hsize37truepc\vsize61truepc
\hoffset=-.5truein\voffset=-0.8truein
\setlength{\baselineskip}{17pt plus 1pt minus 1pt}
\setlength{\textheight}{22.5cm}

\def\diag{{\rm diag}}
\def\I{{\rm i}}
\def\tr{{\rm tr}}
\def\di{{\rm d}}
\def\boldsp{\mbox{\boldmath $\omega$}^A}
\def\boldalpha{\mbox{\boldmath $\alpha$}}
\def\boldbeta{\mbox{\boldmath $\beta$}}
\def\smallboldalpha{\mbox{\small \boldmath $\alpha$}}
\def\smallboldbeta{\mbox{\small \boldmath $\beta$}}
\def\boldomega{\mbox{\boldmath $\omega$}}
\def\rlx{\relax\leavevmode}
\def\inbar{\vrule height1.5ex width.4pt depth0pt}
\def\IC{\rlx\hbox{\,$\inbar\kern-.3em{\rm C}$}}
\def\smallfrac#1#2{\mbox{\small $\frac{#1}{#2}$}}

\input epsf

\begin{titlepage}

\noindent
April, 1995 \hfill{MRR 036-95}\\
\mbox{ } \hfill{hep-th/9506074}
\vskip 1.6in
\begin{center}
{\Large {\bf $O(n)$ model on the honeycomb lattice via }}\\[8pt]
{\Large {\bf reflection matrices: Surface critical behaviour}}
\end{center}

\normalsize
\vskip .3in

\begin{center}
C. M. Yung\footnote{Address after June 1, 1995: Research Institute of
Mathematical Sciences, Kyoto University, Kyoto 606, Japan}   \hspace{3pt}
and \hspace{3pt} M. T. Batchelor
\par \vskip .1in \noindent
{\it Department of Mathematics, School of Mathematical Sciences}\\
{\it Australian National University, Canberra ACT 0200, Australia}
\end{center}
\par \vskip .25in

\begin{center}
{\Large {\bf Abstract}}\\
\end{center}

We study the $O(n)$ loop model on the honeycomb lattice with open boundary
conditions. Reflection matrices for the underlying Izergin-Korepin $R$-matrix
lead to three inequivalent sets of integrable boundary weights. One set,
which has previously been considered, gives rise to the ordinary surface
transition. The other two sets correspond respectively to the special surface
transition and the mixed ordinary-special transition. We analyse the Bethe
ansatz equations derived for these integrable cases and obtain the surface
energies together with the central charges and scaling dimensions
characterizing the corresponding phase transitions.

\vspace{1cm}

\end{titlepage}

\section{Introduction}
\setcounter{equation}{0}

The $O(n)$ model \cite{Stanley68} in two dimensions
has attracted considerable attention in recent years
(see, e.g.\ Refs.\ [2-21] and references therein),
not least because of its connection to the problem of planar
self-avoiding walks or polymers in the $n\rightarrow 0$ limit
\cite{deGennes79}. Methods used to study the model have
included, among others, Coulomb gas techniques
\cite{Nienhuis82,Nienhuis87,Duplantier89}, conformal invariance
\cite{Cardy84,Duplantier86,Cardy87,Burkhardt87,Burkhardt89,Burkhardt94},
Bethe ansatz \cite{Baxter86,Batchelor88,Suzuki88,Suzuki92,Batchelor93},
bootstrap $S$-matrix \cite{Zamolodchikov91,Fendley93,Cardy93,Fendley94}
and numerical transfer matrix calculations \cite{Blote89}.
The self-avoiding walk problem has also been extensively studied
via a number of techniques (see, e.g.\ \cite{DeBell93} and references
therein). Of particular relevance here are the finite-size transfer matrix
calculations \cite{Guim89,Guim94}.

The critical behaviour of the $O(n)$ model is well understood,
with an exact Bethe ansatz solution on the periodic
honeycomb lattice \cite{Baxter86,Batchelor88,Suzuki88} providing a
confirmation of the earlier Coulomb gas and conformal invariance based
results in the {\em bulk}. However, at least from the perspective of
exactly solved lattice models, the situation is not as satisfactory for
{\em surface} critical behaviour. A Bethe ansatz solution on the open
honeycomb lattice was found \cite{Batchelor93}, again confirming earlier
results \cite{Cardy84,Duplantier86,DeBell93} for the ordinary
surface transition \cite{Binder83}.  However, predictions also
exist \cite{Guim89,Burkhardt89,Fendley94,Burkhardt94} for other
surface transitions \cite{Binder83}.
In this paper we derive exact results for the special surface transition,
also on the honeycomb lattice,
using a Bethe ansatz technique which utilises reflection matrices in an
essential way \cite{Yung95a}. We also derive exact results for the model with
mixed ordinary and special boundary conditions. These results for the surface
critical behaviour both complement and, in some cases, extend those currently
available in the literature from other means. Our results for the special
transition and its relevance to the problem of the adsorption transition for
polymers were announced in \cite{Batchelor95a}.

\begin{figure}[htb]
\epsfxsize = 7cm
\vbox{\vskip .8cm\hbox{\centerline{\epsffile{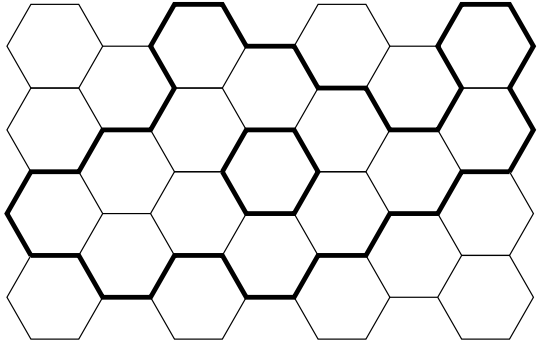}}}
\vskip .5cm \smallskip}
\caption{Honeycomb lattice of ``width'' $N=8$ on which the $O(n)$ model is
defined. The lattice extends to infinity in both vertical directions.
A typical configuration ${\cal G}$ contributing to the partition sum
is shown in bold, with $P=2$, $N_l=2$, $N_r=4$ and $N_b=36$.
}
\end{figure}

The $O(n)$ or $n$-vector model on the honeycomb lattice is a statistical
mechanical model of $n$-dimensional unit vectors $\vec{S}_i$ living on the
sites $i$ of the lattice and interacting via nearest neighbour couplings. Its
partition function is given by
\begin{equation}
Z' = \int \left(\prod_i \di \vec{S}_i \right) \exp \left(
  -\sum_{\langle ij \rangle}
   J_{ij} \vec{S}_i \cdot \vec{S}_j\right).
\end{equation}
On an infinite lattice the coupling constant $J_{ij}$ is usually chosen to
assume the same value $J$ for all nearest neighbour pairs $\langle ij
\rangle$. In this paper we study the model with open boundary conditions
(see Fig 1) with $J_{ij}$ taking values $J_l$ (respectively, $J_r$) if the
nearest neighbour pair $\langle ij\rangle$ lives on the left (right)
boundary. Following Domany et.\ al.\ \cite{Domany81},
we argue that its critical behaviour is
in common with that of a more amenable (but still $O(n)$-symmetric) model
whose partition function is given by
\begin{equation}
Z = \int \left(\prod_i \di \vec{S}_i \right) \prod_{\langle ij \rangle}
  \left( 1+ \kappa_{ij}  \vec{S}_i \cdot \vec{S}_j \right).
\end{equation}
Here the nearest neighbour coupling constant $\kappa_{ij}$ takes values
$\kappa_l$ (respectively, $\kappa_r$) if the pair $\langle ij\rangle$
lives on the left (right) boundary, and $\kappa$ otherwise.

Through a
``high-temperature expansion'' (in small $\kappa$, $\kappa_l$ and $\kappa_r$)
$Z$ can be equivalently written as the partition sum
\begin{equation}
Z = \sum_{{\cal G}} \kappa^{N_b} \kappa_l^{N_l} \kappa_r^{N_r} n^P,
\label{eqn:psum}
\end{equation}
over all configurations ${\cal G}$ of $P$ non-intersecting and closed
loops of fugacity
$n$ on the same lattice,
with $N_l$ (respectively, $N_r$) bonds of ${\cal G}$ living on the
left (right) boundary and $N_b$ bonds not living on either boundary.
When $n\rightarrow 0$ only the empty graph contributes to the partition
function. However, the $\ell$ lines joining the two points $\vec{x}$ and
$\vec{y}$ in a ``watermelon correlator'' $G_{\ell}(\vec{x}-\vec{y})$
survives, giving rise in the continuum limit to the correlation function
of $\ell$ self- and mutually avoiding walks tied at the ends.

The partition sum (\ref{eqn:psum}) can also be
written in the following way which will eventually allow a mapping of the
(non-local) loop model onto a (local)
vertex model:
\begin{equation}
Z \sim \sum_{{\cal G}} w_1^{N_1}\cdots w_{12}^{N_{12}} n^P,
\label{eqn:psum2}
\end{equation}
where $w_i$ are Boltzmann weights for the allowed vertices depicted in
Figure 2, of which there occurs $N_i$ such in the configuration ${\cal G}$.
The weights $w_i$ are given by $w_1=\cdots=w_6=1$, $w_7=w_8=\kappa^{-1}
\equiv t$, $w_9=w_{10}=1$, $w_{11}=\kappa \kappa_l^{-2} \equiv t_l$
and $w_{12}=\kappa \kappa_r^{-2} \equiv t_r$. The partition sum
(\ref{eqn:psum2}) is defined up to an unimportant overall multiplicative
factor.

\begin{figure}[htb]
\epsfxsize = 9cm
\vbox{\vskip .8cm\hbox{\centerline{\epsffile{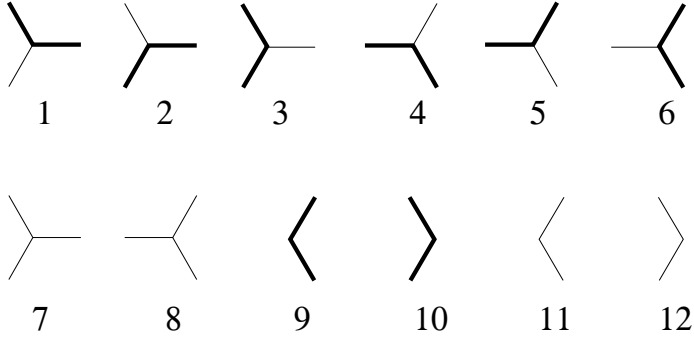}}}
\vskip .5cm \smallskip}
\caption{Allowed loop vertices for the honeycomb loop model. Vertex $i$
carries a Boltzmann weight $w_i$.}
\end{figure}

For the $O(n)$ model on the (infinite) honeycomb lattice, the critical
line
\begin{equation}
\kappa^{-1} = \left(2+(2-n)^{1/2}\right)^{1/2}
\label{eqn:mani}
\end{equation}
was first obtained by Nienhuis \cite{Nienhuis82} using a mapping
to various models and the Coulomb gas.
Baxter \cite{Baxter86} subsequently showed that a vertex
model to which the loop model (\ref{eqn:psum}) can be mapped is solvable
by the coordinate Bethe ansatz precisely along the line
(\ref{eqn:mani}). The Bethe
ansatz solution was then used to derive critical properties for the $O(n)$
model in Refs.\ \cite{Batchelor88,Suzuki88}. For instance, in the
parametrization
\begin{eqnarray}
n &=& -2 \cos \pi g,\\
\kappa^{-1}= t &=& 2 \sin \frac{\pi g}{4},
\end{eqnarray}
the model has central charge
\begin{equation}
c = 1 - \frac{6(1-g)^2}{g}.
\label{eqn:conformal}
\end{equation}
For the open honeycomb lattice
of relevance to this paper, Batchelor and Suzuki \cite{Batchelor93} showed
that along the line specified by (\ref{eqn:mani}) and, in addition,
$\kappa_l=\kappa_r=\kappa$, the model is likewise solvable by the coordinate
Bethe ansatz and corresponds to the $O(n)$ model at the ordinary surface
transition. The Bethe ansatz analysis showed that the central charge
(\ref{eqn:conformal}) remains valid. Furthermore, the model has the set
of scaling dimensions
\begin{equation}
X_{\ell}^{\rm O-O} = h_{\ell+1,1},
\label{eqn:xoo}
\end{equation}
in agreement with an earlier identification in terms of the
Kac formula, where \cite{Duplantier86}
\begin{equation}
h_{p,q} = \smallfrac{1}{4} g p^2 - \smallfrac{1}{2} pq + \frac{q^2 -(g-1)^2}
  {4g}.
\label{eqn:kac}
\end{equation}
Physically, this set of geometric dimensions characterizes the algebraic
decay of the half-watermelon correlators $G_{\ell}(\vec{x}-\vec{y})$ at
criticality \cite{Duplantier86}.

The present authors have recently presented a systematic approach
\cite{Yung95a} to obtain integrable loop models with open boundaries. This
approach begins with Sklyanin's extension of the Quantum Inverse
Scattering Method to obtain integrable open spin chains
\cite{Sklyanin88,Mezincescu91a}. By incorporating
inhomogeneities and making them alternate with the spectral parameter
\cite{Destri92} the Sklyanin transfer matrix becomes relevant to
vertex models with open boundaries. For certain choices of $K$-matrices,
or solutions to the reflection equation, as input to the Sklyanin transfer
matrix the corresponding vertex model can be mapped to a loop model.

The $O(n)$ model of interest here can be obtained from a special case of the
Izergin-Korepin \cite{Izergin81} or $A_2^{(2)}$ vertex model
\cite{Nienhuis90,Reshetikhin91} (see also the reference to Korepin in Ref.\
\cite{Baxter86}). There are three inequivalent diagonal $K$-matrices
associated to the $A_2^{(2)}$ $R$-matrix \cite{Mezincescu91b}. Of these
the two giving rise to {\em non}-quantum group invariant spin chains
allow a mapping to loop models. We showed in \cite{Yung95a} how to recover
the model of \cite{Batchelor93} which corresponds to the ordinary surface
transition. In addition we also found a different set of boundary weights for
which the model is integrable; these correspond to
\begin{equation}
\kappa_l=\kappa_r= \left(-2\cos \smallfrac{\pi g}{2} \right)^{-1/2}
\equiv \kappa_{\rm Sp}.
\label{eqn:klr}
\end{equation}
Note that this critical coupling is related to $n$ through
\begin{equation}
\kappa_{Sp}=(2-n)^{-1/4}.
\label{eqn:spmani}
\end{equation}
We shall presently argue that this solution
corresponds to the $O(n)$ model at the special transition, with central
charge (\ref{eqn:conformal}) and the set of scaling dimensions
\begin{equation}
X_{\ell}^{\rm Sp-Sp} = h_{\ell+1,3},
\label{eqn:xss}
\end{equation}
confirming a conjecture using bootstrap boundary $S$-matrices \cite{Fendley94}.
We also obtain a third set of integrable boundary weights which is a mixture of
the first two; namely,
\begin{equation}
\kappa_l = \kappa, \hspace{20pt} \kappa_r = \kappa_{\rm Sp}.
\label{eqn:kosp}
\end{equation}
We argue that this corresponds to the $O(n)$ model with
mixed ordinary-special boundary conditions, with central charge
\begin{equation}
c^{\rm O-Sp} = 1 - \frac{6(2-g)^2}{g},
\label{eqn:ccosp}
\end{equation}
or equivalently $c^{\rm O-Sp} = c - 24 \;t^{\rm O-Sp}$ with the
``mixed-boundary''
scaling index given by $t^{\rm O-Sp}=h_{1,2}$,
confirming a recent result obtained
using conformal invariance \cite{Burkhardt94}.
In addition, we find the set of scaling dimensions
\begin{equation}
X_{\ell}^{\rm O-Sp}  = h_{\ell+1,1}- \smallfrac{1}{2} \ell.
\label{eqn:xosp}
\end{equation}
A brief summary of our results for mixed boundaries and their
relevance to two-dimensional polymers has been given
elsewhere \cite{Batchelor95b}.

\section{Honeycomb limit of the $A^{(2)}_2$ loop model}
\setcounter{equation}{0}

In this section we present the derivation of integrable $O(n)$ loop models on
the honeycomb
lattice with open boundaries from the $A^{(2)}_2$ $R$-matrix and associated
diagonal $K$-matrices \cite{Mezincescu91b}. This involves the following steps:
(I) Interpreting the Sklyanin transfer matrix $t(u,\boldomega)$
\cite{Sklyanin88} with alternating inhomogeneities \cite{Destri92}
$\boldomega$ as a
diagonal-to-diagonal transfer matrix $t_{\rm D}(u)$ for a vertex model on the
square lattice with open boundaries. (II) Mapping the vertex model to a
loop model on the same lattice. (III) Taking the
`the honeycomb limit' to arrive at the honeycomb lattice loop
model and its transfer matrix $t_{\rm H}$. Via the same chain of mappings the
eigenvalue expression for $t(u,\boldomega)$ (and associated Bethe ansatz
equations) gives rise to the corresponding expression for $t_{\rm H}$.
This procedure
has been explained in detail in our previous paper \cite{Yung95a} for several
three-state $R$-matrices. In particular, the $O(n)$ models with non-mixed
boundaries to which we hitherto refer as O-O and Sp-Sp are treated therein.
We therefore concentrate here on the mixed boundary case
denoted by O-Sp.

\subsection{Sklyanin transfer matrix and eigenvalue expression}

The $R$-matrix for the $A^{(2)}_2$ model \cite{Izergin81} is given
in terms of a spectral parameter $u$
and anisotropy $\lambda$ by
\begin{equation}
R(u)=\left(\begin{array}{ccc|ccc|ccc}
    c & & & & & & & & \\
    & b & & e & & & & &\\
    & & d & & g &  & f & &\\ \hline
    & \bar{e} & & b & & & & &\\
    & & \bar{g}& & a & & g & &\\
    & & & & & b & & e &\\ \hline
    & & & & & & & &\\[-10pt]
    & & \bar{f} & & \bar{g} & & d & &\\
    & & & & & \bar{e} & & b &\\
    & & & & & & & & c \end{array} \right),
\end{equation}
where
\begin{eqnarray*}
a(u) & = & \sinh(u-3\lambda) -
  \sinh(5\lambda) + \sinh(3\lambda) +
   \sinh(\lambda),\\
b(u) & = & \sinh(u-3\lambda) +
  \sinh(3\lambda),\\
c(u) & = & \sinh(u-5\lambda) +
  \sinh(\lambda),\\
d(u) & = & \sinh(u-\lambda) +
  \sinh(\lambda),\\
e(u) & = & -2e^{-\smallfrac{u}{2}}\sinh(2\lambda)
  \cosh(\smallfrac{u}{2}- 3\lambda),\\
\bar{e}(u) & = & -2 e^{\smallfrac{u}{2}}\sinh(2\lambda)
   \cosh(\smallfrac{u}{2}- 3\lambda),\\
f(u) & = & -2 e^{-u+2\lambda}\sinh(\lambda)
  \sinh(2\lambda) - e^{-\lambda}
   \sinh(4\lambda),\\
\bar{f}(u) & = & 2e^{u-2\lambda}\sinh(\lambda)
   \sinh(2\lambda) -
   e^{\lambda}\sinh(4\lambda),\\
g(u) & = & 2 e^{-\smallfrac{u}{2}+
   2\lambda}\sinh(\smallfrac{u}{2})
    \sinh(2\lambda),\\
\bar{g}(u) & = & -2 e^{\smallfrac{u}{2}-2\lambda}
  \sinh(\smallfrac{u}{2})
    \sinh(2\lambda).
\end{eqnarray*}
It was shown in \cite{Mezincescu91b} that there exists three inequivalent
diagonal $K$-matrices $K^-(u)$ satisfying the reflection equation
\begin{eqnarray}
\lefteqn{R_{12}(u-v) \stackrel{1}{K^-}(u)
  R_{21}(u+v)
  \stackrel{2}{K^-}(v)=}\hspace{50pt}\nonumber\\
& & \stackrel{2}{K^-}(v) R_{12}(u+v)
  \stackrel{1}{K^-}(u) R_{21}(u-v).
\label{eqn:k1}
\end{eqnarray}
They are namely, $K^-(u)=1$ and $K^-(u)=\Gamma^{\pm}
(u)$, where
\begin{equation}
\Gamma^{\pm}(u) = \diag( e^{-u} \psi^{\pm}(u),
\phi^{\pm}(u),
  e^{u} \psi^{\pm}(u) ),
\end{equation}
with
\begin{eqnarray}
\psi^{\pm}(u) & = & \cosh(\smallfrac{u}{2}-
  3\lambda) \pm  \I
\sinh(\smallfrac{u}{2})\label{eqn:psi}\\
\phi^{\pm}(u) & = &
\cosh(\smallfrac{u}{2}+3\lambda)\mp \I \sinh
(\smallfrac{u}{2}).
\end{eqnarray}
Let $K^+(u)= K^-(-u-\eta)^t M$,
where $K^-(u)$ is any of the three
$K$-matrices defined above, $\eta=-6\lambda-\I \pi$ is the
crossing parameter
and $M=\diag(e^{2\lambda},1,e^{-2\lambda})$
is the crossing matrix.
Then the Sklyanin transfer matrix \cite{Sklyanin88,Mezincescu91a}
\begin{equation}
 t(u,\boldomega) = \tr_a\; \stackrel{a}{K^+}(u)
 T_a(u,\boldomega)
   \stackrel{a}{K^-}(u) \tilde{T}_a(u,\boldomega),
\label{eqn:skl}
\end{equation}
forms a commuting family: $[t(u,\boldomega),t(v,
\boldomega)]=0$ for arbitrary $u,v$.
Here $T_a$ and $\tilde{T}_a$ are monodromy matrices defined
by
\begin{eqnarray}
 T_a(u,\boldomega) & = & R_{a1}(u+\omega_1)
  R_{a2}(u+\omega_2) \cdots R_{aN}(u+
   \omega_N), \label{eqn:monod1}\\
 \tilde{T}_a(u,\boldomega) & = & R_{Na}(u-\omega_N)
  \cdots R_{2a}(u-\omega_2)R_{1a}(u-\omega_1),
\label{eqn:monod2}
\end{eqnarray}
where $a$ labels the auxiliary space $V_a$ and $1,\ldots,N$ label quantum
spaces. With three choices each for $K^-(u)$ and
$K^+(u)$ we have therefore
nine possiblities for $t(u,\boldomega)$.
The three non-mixed cases were dealt
with in \cite{Yung95a}. It was shown there that the non quantum group-invariant
cases, O-O and Sp-Sp, based on $\Gamma^{\pm}(u)$
give rise to loop models, in
contrast to the quantum group-invariant case with $K^-(u)=1$,
$K^+(u)=M$.
Similar reasoning shows that only two of the mixed cases O-Sp and Sp-O give
rise to loop models. These two loop models are equivalent, as to be expected,
and we henceforth deal only with the case O-Sp specified by
$K^-(u)= \Gamma^-(u)$ and $K^+(u)=
\Gamma^+(-u-\eta)^t M$.

The eigenvalue expression and Bethe ansatz equations for the Sklyanin
transfer matrix $t(u,\boldomega)$
were obtained for the cases O-O and Sp-Sp
in \cite{Yung95a} using what can be referred to as the ``Doubling
Hypothesis'' (cf.\ \cite{Artz95,Yung95b}). For the model at hand, this says
that the eigenvalue expression can be written in the form
\begin{equation}
\Lambda(u,\boldomega) = \sum_{i=1}^3 \alpha_i(u)
\Phi_i(u,\boldomega) A_i(u),
\label{eqn:lamu}
\end{equation}
where only the ``boundary contributions'' $\alpha_i(u)$ vary
with the chosen (diagonal)
$K$-matrices. The functions $\Phi_i(u,\boldomega)$ and $A_i(u)$
are doubled versions of the corresponding functions which appear in the
eigenvalue expression for the transfer matrix with periodic boundary
conditions. For the $A^{(2)}_2$ model they are given by
\begin{eqnarray}
\Phi_1(u,\boldomega) &=& \prod_{i=1}^N c(u+\omega_i)
  c(u-\omega_i),\nonumber\\
\Phi_2(u,\boldomega) &=& \prod_{i=1}^N b(u+\omega_i)
  b(u-\omega_i),\nonumber\\
\Phi_3(u,\boldomega) &=& \prod_{i=1}^N d(u+\omega_i)
  d(u-\omega_i),
\label{eqn:phii}
\end{eqnarray}
where $b(u),c(u),d(u)$ are diagonal matrix elements
of $R(u)$ and
\begin{eqnarray}
A_1(u) & = & \prod_{j=1}^{M} \frac{\sinh[\smallfrac{1}{2}(u
  +u_j) + \lambda]
   \sinh[\smallfrac{1}{2}(u-u_j) +\lambda]}
   {\sinh[\smallfrac{1}{2}(u+u_j) -\lambda]
        \sinh[\smallfrac{1}{2}(u-u_j) -\lambda]},\nonumber\\
A_2(u) & = & \prod_{j=1}^{M} \frac{\sinh[\smallfrac{1}{2}(u
   +u_j) - 3\lambda]
     \sinh[\smallfrac{1}{2}(u-u_j) -3\lambda]}
      {\sinh[\smallfrac{1}{2}(u+u_j) -\lambda]
       \sinh[\smallfrac{1}{2}(u-u_j) -\lambda]}
   \nonumber\label{eqn:b}\\
     &  & \times \; \prod_{j=1}^{M}\frac{\cosh[\smallfrac{1}{2}(u+u_j)]
           \cosh[\smallfrac{1}{2}(u-u_j)]}
       {\cosh[\smallfrac{1}{2}(u+u_j)-2\lambda]
           \cosh[\smallfrac{1}{2}(u-u_j)-
    2\lambda]},\nonumber\\
A_3(u) & = & \prod_{j=1}^{M}\frac{\cosh[\smallfrac{1}{2}(u
   +u_j) - 4\lambda]
   \cosh[\smallfrac{1}{2}(u-u_j) -4\lambda]}
   {\cosh[\smallfrac{1}{2}(u+u_j) -2\lambda]
         \cosh[\smallfrac{1}{2}(u-u_j) -2\lambda]}.
\label{eqn:ai}
\end{eqnarray}
The boundary contributions
$\alpha_i(u)$ are determined by the action of
$t(u,\boldomega)$ on the
reference state $|\Omega\rangle$, being the $N$-fold tensor product of the
vector with $1$ as its first entry and $0$ elsewhere,
and depend on the $K$-matrices $K^{\pm}(u)$ according to \cite{Yung95a}
\begin{eqnarray}
\alpha_1(u) & = & K_{11}^-\left\{ K_{11}^+ + \frac{K_{22}^+ \bar{e}^2}
  {c^2-b^2} + \frac{K_{33}^+}{c^2-d^2}\left( \bar{f}^2 + \frac{\bar{g}^2
  \bar{e}^2}{c^2-b^2}\right)\right\},\nonumber\\
\alpha_2(u) & = & K_{22}^+ K_{22}^- - \frac{K_{22}^+K_{11}^-\bar{e}^2}
  {c^2-b^2}-
   \frac{K_{33}^+K_{22}^-\bar{g}^2}{d^2-b^2} - \frac{K_{33}^+K_{11}^-
 \bar{g}^2\bar{e}^2}{(c^2-b^2)(b^2-d^2)},\nonumber\\
\alpha_3(u) & = & K_{33}^+\left\{ K_{33}^- + \frac{K_{22}^-\bar{g}^2}
  {d^2-b^2} -
  \frac{K_{11}^-}{c^2-d^2}\left( \bar{f}^2 - \frac{\bar{g}^2\bar{e}^2}
  {b^2-d^2}\right) \right\}.
\label{eqn:mnev}
\end{eqnarray}
Here $\bar{e},c$, etc.\ are matrix elements of $R(u)$
whose dependence on $u$,
like those of the matrix elements of $K^{\pm}(u)$,
have been suppressed for convenience.
For O-Sp boundaries we find the following:
\begin{eqnarray}
\alpha_1(u) &=& \frac{\sinh(u-6\lambda)}{c(2u)}
  \psi^+(u) \psi^-(u) \xi^+(u),\\
\alpha_2(u) &=& \frac{\sinh(u-6\lambda)
   \sinh(u)}{c(2u) \sinh(u-4\lambda)}
  \psi^+(u) \psi^-(u) \xi^+(u),\\
\alpha_3(u) &=& \frac{\sinh(u)\sinh(u-
  2\lambda)} {c(2u) \sinh(u-4\lambda)}
  \psi^+(u) \psi^-(u) \xi^+(u),
\end{eqnarray}
where
\begin{equation}
\xi^{\pm}(u) = 2 \left( \cosh(3\lambda-u)
\pm \I \sinh 2 \lambda \right)
\end{equation}
and $\psi^{\pm}(u)$ is as defined in Eq.\ (\ref{eqn:psi}).
Together with Eqs.\ (\ref{eqn:phii}) and (\ref{eqn:ai})
these fully determine the eigenvalue
$\Lambda(u,\boldomega)$ through Eq.\ (\ref{eqn:lamu}).
The corresponding Bethe ansatz equations are obtained
by requiring $\Lambda(u,\boldomega)$ to be analytic in $u$.
It should be pointed out that our justification for the eigenvalue
expression (\ref{eqn:lamu}) is not as strong as we would like, although we
have no doubt about its correctness; a derivation using the algebraic
Bethe ansatz \cite{Sklyanin88} would be preferable. In fact, a derivation
using the analytic Bethe ansatz generalizaling the study of
\cite{Mezincescu92b} for
the quantum group-invariant case would also go a long way towards answering
this
criticism. At present our result is backed up mainly with numerical checks (see
end of Section 2).

\subsection{Vertex and loop models with open boundaries}

\begin{figure}[htb]
\epsfxsize = 9cm
\vbox{\vskip .8cm\hbox{\centerline{\epsffile{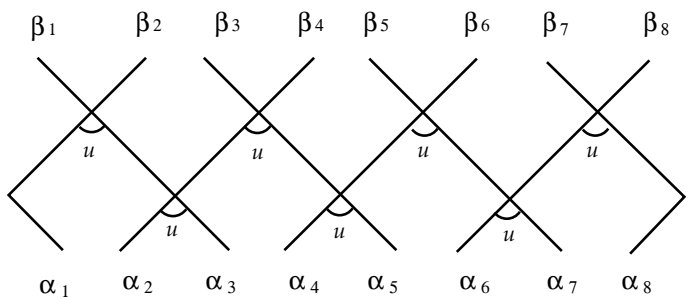}}}
\vskip .5cm \smallskip}
\caption{The ``diagonal-to-diagonal''
transfer matrix $t_D(u)$ which takes
the state $(\alpha_1,\ldots,\alpha_N)$ to the state
$(\beta_1,\ldots,\beta_N)$. All internal edges are summed over.
}
\end{figure}

When the inhomogeneities $\boldomega$ are chosen to alternate as $\omega_i=
(-)^{i+1}u$,
the Sklyanin transfer matrix turns into the diagonal-to-diagonal
transfer matrix $t_{\rm D}(u)$
for a vertex model on the square lattice with
open boundaries \cite{Yung95a} (see also \cite{Destri92}). More specifically,
define the functions
\begin{eqnarray}
  l_a^b(u) & = & \sum_{cd} R_{da}^{bc}(2u)
  K^+_{cd}(u),\nonumber\\
  r_a^b(u) & = & K^-_{ab}(u).
\label{eqn:lr}
\end{eqnarray}
Then the object
\begin{equation}
t_{\rm D}(u) = \frac{t\left(\smallfrac{u}{2},
\omega_j=(-)^{j+1} \smallfrac{u}{2} \right) R_{11}^{11}(u)}
 {\Phi_1\left(\smallfrac{u}{2},\omega_j=(-)^{j+1}
  \smallfrac{u}{2} \right)l^1_1(\smallfrac{u}{2})
  r^1_1(\smallfrac{u}{2})},
\end{equation}
is the diagonal-to-diagonal transfer matrix (see Fig. 3) for the
vertex model with bulk weights given by
\begin{center}
$$
\setlength{\unitlength}{0.012500in}%
\begingroup\makeatletter
\def\x#1#2#3#4#5#6#7\relax{\def\x{#1#2#3#4#5#6}}%
\expandafter\x\fmtname xxxxxx\relax \def\y{splain}%
\ifx\x\y   
\gdef\SetFigFont#1#2#3{%
  \ifnum #1<17\tiny\else \ifnum #1<20\small\else
  \ifnum #1<24\normalsize\else \ifnum #1<29\large\else
  \ifnum #1<34\Large\else \ifnum #1<41\LARGE\else
     \huge\fi\fi\fi\fi\fi\fi
  \csname #3\endcsname}%
\else
\gdef\SetFigFont#1#2#3{\begingroup
  \count@#1\relax \ifnum 25<\count@\count@25\fi
  \def\x{\endgroup\@setsize\SetFigFont{#2pt}}%
  \expandafter\x
    \csname \romannumeral\the\count@ pt\expandafter\endcsname
    \csname @\romannumeral\the\count@ pt\endcsname
  \csname #3\endcsname}%
\fi
\endgroup
\begin{picture}(60,64)(75,650)
\thicklines
\put(100,675){\oval( 10, 10)[bl]}
\put(100,675){\oval( 10, 10)[br]}
\put( 80,700){\line( 1,-1){ 40}}
\put(120,700){\line(-1,-1){ 40}}
\put( 95,660){\makebox(0,0)[lb]{\smash{\SetFigFont{12}{14.4}{it}u}}}
\put( 72,650){\makebox(0,0)[lb]{\smash{\SetFigFont{12}{14.4}{it}i}}}
\put(120,705){\makebox(0,0)[lb]{\smash{\SetFigFont{12}{14.4}{it}k}}}
\put( 75,705){\makebox(0,0)[lb]{\smash{\SetFigFont{12}{14.4}{it}l}}}
\put(120,650){\makebox(0,0)[lb]{\smash{\SetFigFont{12}{14.4}{it}j}}}
\put(135,675){\makebox(0,0)[lb]{\smash{\SetFigFont{12}{14.4}{rm}=}}}
\put(155,675){\makebox(0,0)[lb]{\smash{\SetFigFont{17}{20.4}{rm}
$\frac{R_{i j}^{k l}(u)}{R_{11}^{11}(u)}$}}}
\end{picture}
$$
\end{center}
and boundary weights given by
$$
\setlength{\unitlength}{0.012500in}%
\begingroup\makeatletter
\def\x#1#2#3#4#5#6#7\relax{\def\x{#1#2#3#4#5#6}}%
\expandafter\x\fmtname xxxxxx\relax \def\y{splain}%
\ifx\x\y   
\gdef\SetFigFont#1#2#3{%
  \ifnum #1<17\tiny\else \ifnum #1<20\small\else
  \ifnum #1<24\normalsize\else \ifnum #1<29\large\else
  \ifnum #1<34\Large\else \ifnum #1<41\LARGE\else
     \huge\fi\fi\fi\fi\fi\fi
  \csname #3\endcsname}%
\else
\gdef\SetFigFont#1#2#3{\begingroup
  \count@#1\relax \ifnum 25<\count@\count@25\fi
  \def\x{\endgroup\@setsize\SetFigFont{#2pt}}%
  \expandafter\x
    \csname \romannumeral\the\count@ pt\expandafter\endcsname
    \csname @\romannumeral\the\count@ pt\endcsname
  \csname #3\endcsname}%
\fi
\endgroup
\begin{picture}(180,57)(20,755)
\thicklines
\put( 40,800){\line(-1,-1){ 20}}
\put( 20,780){\line( 0, 1){  0}}
\put( 20,780){\line( 1,-1){ 20}}
\put(160,800){\line( 1,-1){ 20}}
\put(180,780){\line( 0, 1){  0}}
\put(180,780){\line(-1,-1){ 20}}
\put( 45,755){\makebox(0,0)[lb]{\smash{\SetFigFont{12}{14.4}{it}i}}}
\put( 45,800){\makebox(0,0)[lb]{\smash{\SetFigFont{12}{14.4}{it}j}}}
\put( 55,775){\makebox(0,0)[lb]{\smash{\SetFigFont{12}{14.4}{rm}=}}}
\put( 75,775){\makebox(0,0)[lb]{\smash{\SetFigFont{17}{20.4}{rm}
$\frac{l_i^j(u/2)}{l_1^1(u/2)}$,}}}
\put(150,755){\makebox(0,0)[lb]{\smash{\SetFigFont{12}{14.4}{it}i}}}
\put(150,800){\makebox(0,0)[lb]{\smash{\SetFigFont{12}{14.4}{it}j}}}
\put(195,775){\makebox(0,0)[lb]{\smash{\SetFigFont{12}{14.4}{rm}=}}}
\put(215,775){\makebox(0,0)[lb]{\smash{\SetFigFont{17}{20.4}{rm}
$\frac{r_i^j(u/2)}{r_1^1(u/2)}$.}}}
\end{picture}
$$
For alternating
inhomogeneities $\omega_i=(-)^{i+1}u$
we note that $\Phi_2(u,\boldomega)$
and $\Phi_3(u,\boldomega)$
vanish due to $b(0)=d(0)=0$. Using the identity $\alpha_1(u)=
l^1_1(u) r^1_1(u) / R_{11}^{11}(u)$
we arrive at the eigenvalue expression
\begin{equation}
t_{\rm D}(u) = A_1(\smallfrac{u}{2}),
\label{eqn:td}
\end{equation}
with the function $A_1(u)$ defined in (\ref{eqn:ai}).
The associated Bethe ansatz equations
are those which render $\Lambda(u,\boldomega)$ analytic in $u$,
but with the inhomogeneities subsequently specialized to
$\omega_j=(-)^{j+1}u$. Note
that the order is crucial; analyticity of $t_{\rm D}(u)$
does not give the Bethe ansatz equations. With this we accomplish step (I).

We shall presently make a change of variables
$u\rightarrow 2\I u, \lambda\rightarrow \I \lambda+ \I \pi/2$,
which takes us into the critical regime of
the model for real $u,\lambda$. The eigenvalue expression
for $t_{\rm D}(u)$ becomes (after a shift $u_j \rightarrow u_j + \I\pi$)
\begin{equation}
\Lambda_{\rm D}(u) =
\prod_{j=1}^M \frac{\sinh[\smallfrac{1}{2}(\I u + u_j) +
   \I \lambda]
             \sinh[\smallfrac{1}{2}(\I u - u_j) + \I \lambda]}
            {\sinh[\smallfrac{1}{2}(\I u + u_j) -  \I \lambda]
             \sinh[\smallfrac{1}{2}(\I u - u_j)  - \I
\lambda]}
\label{eqn:c2ev}
\end{equation}
while the Bethe ansatz equations now take the form ($k=1,\ldots,M$)
\begin{eqnarray}
\lefteqn{\frac{\sinh(\I \lambda-u_k)}{\sinh(\I \lambda + u_k)}
 \left(\frac{\sinh[\smallfrac{1}{2}(u_k - 2 \I \lambda-\I u)]
           \sinh[\smallfrac{1}{2}(u_k -  2\I \lambda + \I u)]}
           {\sinh[\smallfrac{1}{2}(u_k + 2 \I \lambda+ \I u)]
           \sinh[\smallfrac{1}{2}(u_k +  2 \I \lambda-\I u)]}
\right)^N} \hspace{70pt} \nonumber\\
&= & \prod_{j\neq k}^{M} \frac{\sinh[\smallfrac{1}{2}(u_k+u_j)-2 \I
    \lambda]
 \sinh[\smallfrac{1}{2}(u_k-u_j)-2\I \lambda]}
{\sinh[\smallfrac{1}{2}(u_k+u_j)+2\I \lambda]
 \sinh[\smallfrac{1}{2}(u_k-u_j)+2\I \lambda]} \nonumber\\
& & \times \; \frac{
\sinh[\smallfrac{1}{2}(u_k+u_j)+\I \lambda]
 \sinh[\smallfrac{1}{2}(u_k-u_j)+\I \lambda]}
{\sinh[\smallfrac{1}{2}(u_k+u_j)-\I \lambda]
 \sinh[\smallfrac{1}{2}(u_k-u_j)-\I \lambda]}.
\label{eqn:c2bae}
\end{eqnarray}

\begin{figure}[htb]
\epsfxsize = 12cm
\vbox{\vskip .8cm\hbox{\centerline{\epsffile{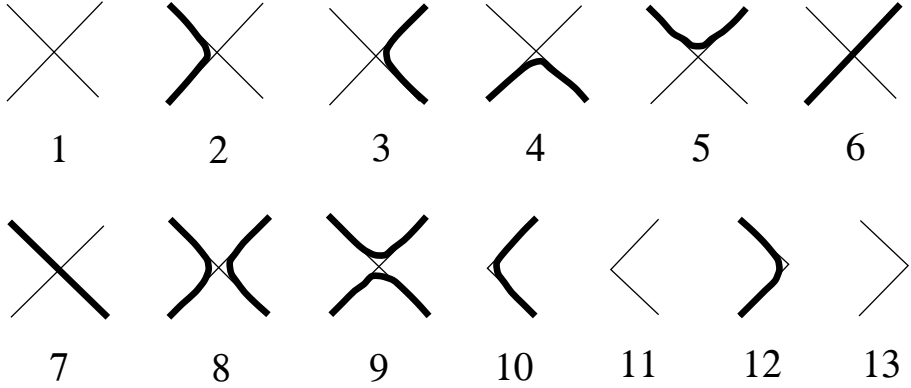}}}
\vskip .5cm \smallskip}
\caption{Allowed vertices for the loop model with partition function
$Z_{\rm loop}$. Vertex $i$ carries Boltzmann weight $\rho_i$.}
\end{figure}

Using an argument very similar to that for O-O and Sp-Sp \cite{Yung95a} one
can show that the vertex model partition function is equivalent to that
of a loop model on the same lattice
\begin{equation}
Z_{\rm loop} = \sum_{{\cal G}} \rho_1^{N_1} \cdots \rho_{13}^{N_{13}} \; n^{P},
\end{equation}
where the sum is over all configurations ${\cal G}$ of non-intersecting closed
loops which cover some (or none) of the edges, with non-zero
Boltzmann weights $\rho_i$ shown in Figure 4. In the configuration ${\cal G}$,
$N_i$ is the number of occurences of the weight of type $i$ while $P$ is the
total number of closed loops of fugacity $n$. The explicit expressions for
$\rho_i$ and $n$ are given (in the shifted variables $u,\lambda$) by
\begin{eqnarray}
\rho_1 & = & \left[\sin(3\lambda-u)\sin(u) +
       \sin(2\lambda)\sin(3\lambda)\right]
       \left[\sin(3\lambda-
      u) \sin(2\lambda-u)\right]^{-1},\nonumber\\
\rho_2 = \rho_3 & = &\epsilon_1 \sin(2\lambda)
      \sin(2\lambda-u)^{-1},\nonumber\\
\rho_4 = \rho_5 & = & \epsilon_2 \sin(u)\sin(2\lambda)
      \left[\sin(3\lambda-u) \sin(2\lambda-u)
      \right]^{-1},
    \nonumber\\
\rho_6 = \rho_7 & = & \sin(u)
    \sin(2\lambda-u)^{-1}, \nonumber\\
\rho_8 &=& 1,\nonumber\\
\rho_9 &=& - \sin(\lambda-u)\sin(u)
     \left[\sin(3\lambda-u) \sin(2\lambda-u)
     \right]^{-1},
     \nonumber\\
\rho_{10} & = & \epsilon_1 e^{-\I u},\nonumber\\
\rho_{11} & = &  e^{-\I u} \sin[\smallfrac{1}{2}(
   3\lambda+u) ] \sin[\smallfrac{1}{2}(
   3\lambda- u)]^{-1},\nonumber\\
\rho_{12} & = & \epsilon_1 e^{\I u},\nonumber\\
\rho_{13} & = &  e^{\I u} \cos[\smallfrac{1}{2}(
   3\lambda+u) ] \cos[\smallfrac{1}{2}(
   3\lambda- u)]^{-1},\nonumber\\
n & = & -2\cos(4\lambda)
\label{eqn:lbulk}
\end{eqnarray}
where $\epsilon_1^2=\epsilon_2^2=1$. This concludes step (II).

To accomplish step (III) we now choose $\epsilon_1=\epsilon_2=1$ and
take the ``honeycomb limit''
$u=\lambda$. As seen from Eqs.\ (\ref{eqn:lbulk}),
the Boltzmann weight $\rho_9$ vanishes in this limit allowing the
remaining bulk vertices to be ``pulled apart horizontally'' to give the
vertices of Figure 5. It is apparent that these latter vertices
define a honeycomb lattice loop model. In other words, in the honeycomb
limit the square lattice loop model defined through the weights of Figure 4
is equivalent to the honeycomb lattice loop model defined through the
weights of Figure 5. There is in fact a second honeycomb limit, which we
will consider no further, where $\rho_8$ vanishes and the remaining bulk
vertices are pulled apart vertically.

The left and right boundary weights can be normalized separately such that
$\rho_{10}=\rho_{12}=1$, while keeping $Z_{\rm loop}$ unchanged. In the
honeycomb limit the bulk
weights are given by $\rho_4=\rho_5=\rho_6=\rho_7=\rho_8=1$, $\rho_2=\rho_3=
2\cos(\lambda)\equiv t$ and $\rho_1=t^2$.
Define the weights
\begin{equation}
t_{\rm O}\equiv t=2\cos(\lambda),\hspace{40pt}
t_{\rm Sp} \equiv \cos(2 \lambda) \cos(\lambda)^{-1}.
\end{equation}
{F}rom Eqs.\ (\ref{eqn:lbulk}) we see that the remaining boundary
weights are $\rho_{11}=t_{\rm O}$ and $\rho_{13}=t_{\rm Sp}$. An inspection
of Figure 5 shows that we can achieve the same effect of the weight assignment
$\rho_i$ by letting a
honeycomb vertex (bulk or boundary) through which a bond runs carry a
weight $1$, an ``empty'' bulk vertex  a weight $t$, an empty left
boundary vertex a weight $t_{\rm O}$ and an empty right
boundary vertex a weight $t_{\rm Sp}$. This justifies the weight assignment
(\ref{eqn:kosp}) in the context of Figure 2, after a change of variables from
$\lambda$ to $g$ related by
\begin{equation}
\pi g = 2\pi -4\lambda.
\label{eqn:lg}
\end{equation}

\begin{figure}[htb]
\epsfxsize = 12cm
\vbox{\vskip .8cm\hbox{\centerline{\epsffile{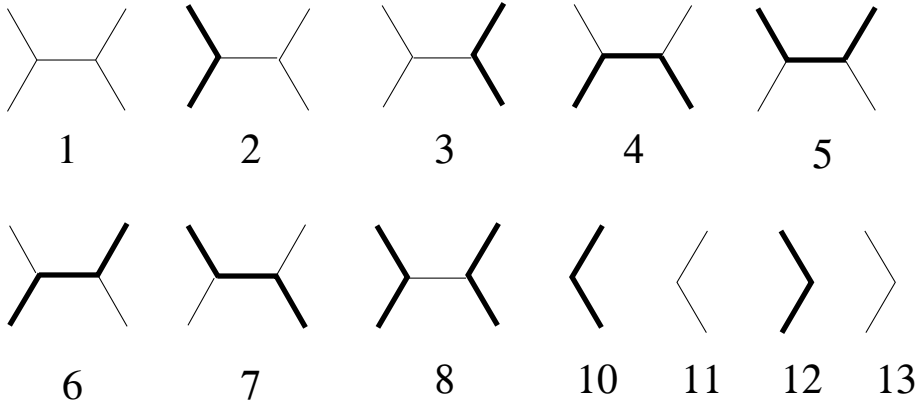}}}
\vskip .5cm \smallskip}
\caption{The vertices of Figure 4 after the honeycomb limit has been
taken.}
\end{figure}

In the honeycomb limit, the diagonal-to-diagonal transfer matrix $t_{\rm D}(u)$
of Figure 3 becomes the transfer matrix $t_{\rm H}$ for taking a row of
edges up the honeycomb strip of Figure 1 to the next similar row.
The eigenvalue expression (\ref{eqn:td}) then becomes (with a rescaling
$u_k \rightarrow 2 u_k$)
\begin{equation}
\Lambda_{\rm H} =
\prod_{j=1}^M \frac{\sinh(u_j + \smallfrac{3}{2} \I \lambda)
   \sinh(u_j - \smallfrac{3}{2} \I \lambda)}
   {\sinh(u_j + \smallfrac{1}{2} \I \lambda)
   \sinh(u_j - \smallfrac{1}{2} \I \lambda)}
\label{eqn:honeyev}
\end{equation}
The associated Bethe ansatz equations become ($k=1,\ldots,M$)
\begin{eqnarray}
\lefteqn{\Theta_{\rm O-Sp} (u_k,\lambda)
 \left[\frac{\sinh(u_k-\smallfrac{1}{2} \I \lambda)
    \sinh(u_k-\smallfrac{3}{2} \I \lambda)}
    {\sinh(u_k+\smallfrac{1}{2} \I \lambda)
    \sinh(u_k+\smallfrac{3}{2} \I \lambda)}
\right]^N} \hspace{70pt} \nonumber\\
&= & \prod_{j\neq k}^{M} \frac{\sinh(u_k+u_j-2 \I \lambda)
 \sinh(u_k-u_j-2\I \lambda)}
 {\sinh(u_k+u_j+2 \I \lambda)\sinh(u_k-u_j-2\I \lambda)}\nonumber\\
& & \times \; \frac{\sinh(u_k+u_j+ \I \lambda)
 \sinh(u_k-u_j+\I \lambda)}
 {\sinh(u_k+u_j-\I \lambda)\sinh(u_k-u_j-\I \lambda)},
\label{eqn:honeybae}
\end{eqnarray}
with the ``boundary function'' $\Theta_{\rm O-Sp}(u,\lambda)$ defined as
\begin{equation}
\Theta_{\rm O-Sp}(u,\lambda)=\frac{\sinh(\I \lambda - 2u)}
 {\sinh(\I \lambda + 2u)}.
\label{eqn:Thos}
\end{equation}
In the language of Figure 2, the boundary weights for the O-Sp model are
given by
\begin{equation}
t_l = t_{\rm O}, \hspace{40pt} t_r = t_{\rm Sp}.
\label{eqn:tos}
\end{equation}
The corresponding weights for the O-O model \cite{Batchelor93,Yung95a} are
\begin{equation}
t_l = t_{\rm O}, \hspace{40pt} t_r = t_{\rm O}.
\label{eqn:too}
\end{equation}
The transfer matrix eigenvalue expression (\ref{eqn:honeyev}) remains
valid while the Bethe
ansatz equations (\ref{eqn:honeybae}) have the boundary function
$\Theta_{\rm O-Sp}(u,\lambda)$ replaced by
\begin{equation}
\Theta_{\rm O-O}(u,\lambda)= \left[\frac{\sinh(u-\smallfrac{1}{2}\I \lambda)}
  {\sinh(u+\smallfrac{1}{2}\I \lambda)}\right]^2.
\label{eqn:Thoo}
\end{equation}
For the Sp-Sp model \cite{Yung95a} the corresponding weights are
\begin{equation}
t_l = t_{\rm Sp}, \hspace{40pt} t_r = t_{\rm Sp},
\label{eqn:tss}
\end{equation}
again with transfer matrix eigenvalue expression (\ref{eqn:honeyev}) and Bethe
ansatz equations (\ref{eqn:honeybae}) with boundary function
\begin{equation}
\Theta_{\rm Sp-Sp}(u,\lambda)=
  \left[\frac{\cosh(u-\smallfrac{1}{2}\I \lambda)}
  {\cosh(u+\smallfrac{1}{2}\I \lambda)}\right]^2
\label{eqn:Thss}
\end{equation}
in place of $\Theta_{\rm O-Sp}(u,\lambda)$.

Note the relation
\begin{equation}
\Theta_{\rm O-Sp}(u,\lambda)^2 =\Theta_{\rm O-O}(u,\lambda)
\Theta_{\rm Sp-Sp}(u,\lambda)
\label{eqn:relth}
\end{equation}
between the various boundary functions. The choice of boundary weights
-- (\ref{eqn:tos}), (\ref{eqn:too}) or (\ref{eqn:tss}) -- and corresponding
boundary function -- (\ref{eqn:Thos}), (\ref{eqn:Thoo}) or
(\ref{eqn:Thss}) -- which arise from the appropriate combination of
$K$-matrices governs the surface critical behaviour of the model. In the
next section both universal and non-universal quantities will be derived
from the eigenvalue expression (\ref{eqn:honeyev}) and Bethe ansatz equations
(\ref{eqn:honeybae}) and their analogues for O-O and Sp-Sp boundaries.

We first solved the Bethe ansatz equations numerically and compared the
eigenvalues so obtained to those from an exact numerical
diagonalization of the transfer matrix for the vertex model (and also the
loop model). The objective was twofold; firstly to confirm that the results
obtained using the Doubling Hypothesis (\ref{eqn:lamu}) are correct, and
secondly to obtain the Bethe ansatz root distributions corresponding to
various eigenstates of interest.

Restricting ourselves to even lattice sizes $N$,
our chief findings can be summarised thus: Let
$\ell=N-M$ label the sectors of the (vertex model) transfer matrix and
$\Lambda_{\ell}$ be the leading eigenvalue in that sector. Then $\ell$ takes
on values $0,1,\ldots,N-1$. For instance,
$\Lambda_0$ is the largest eigenvalue in the ground state sector. For all
cases except $\ell=0$ for Sp-Sp boundaries, the eigenvalue $\Lambda_{\ell}$
corresponds to {\em all} $N-\ell$ Bethe ansatz roots being real and strictly
positive. For the
dominant eigenvalue $\Lambda_0^{\rm Sp-Sp}$ there are $N-1$ positive
roots and
a pure imaginary root at approximately $u_1 \approx \I\left(\smallfrac{\pi}{4}
+ \smallfrac{\lambda}{2}\right)$. The scaling dimensions associated with
$\Lambda_{\ell}$ are argued to give rise to the exponents characterizing
the critical behaviour of the watermelon correlators.

\section{Analysis of the Bethe ansatz equations}
\setcounter{equation}{0}

At criticality, the free energy per site of
the model is expected to scale with lattice size $N$ as
(see, e.g.\ Ref.\ \cite{Burkhardt94} and references therein)
\begin{equation}
f_N = f_{\infty} + \frac{f_S^{(l)}}{N} + \frac{f_S^{(r)}}{N} -
  \frac{\pi \zeta c}{24 N^2} + O(N^{-3}),
\label{eqn:confinv}
\end{equation}
where $c$ is the central charge for the underlying conformal field theory
and here $\zeta$ is a lattice-dependent scale factor.
The bulk free energy
$f_{\infty}$, the surface energies $f_S^{(l)}$ and $f_S^{(r)}$ associated
with the left and right boundaries, and the central charge $c$ are all
calculable from the Bethe ansatz equations, using a technique which is
by now fairly standard
(cf.\ \cite{deVega85,Hamer87,Batchelor88,Suzuki88,Suzuki92,Batchelor93}).
In particular, the calculation for O-O boundaries sketched in
\cite{Batchelor93} is readily adaptable to handle the other two types of
integrable boundaries.

We are mainly interested in the eigenvalues $\Lambda_{\rm H}$ corresponding
to purely real Bethe ansatz roots. The eigenvalue expression
(\ref{eqn:honeybae}) and Bethe ansatz equations (\ref{eqn:logbae}) are
invariant under $\lambda \rightarrow - \lambda$ and $\lambda \rightarrow
\lambda + \pi$.
Hence it is sufficient to consider $\lambda$ in the region $[0,\pi/2]$.
We restrict ourselves to values of $\lambda$ lying in the interval $(0,\pi/2)$
to simplify certain Fourier integrals; the expressions for the central
charges and scaling dimensions are nevertheless valid for the end-points,
as can be shown using a limiting procedure. The eigenvalue
$\Lambda_0^{\rm Sp-Sp}$
will be considered separately towards the end of the paper. Define the function
\begin{equation}
\phi(u,\lambda) = 2 \tan^{-1}\left( \cot \lambda \tanh u \right),
\end{equation}
whose derivative with respect to $u$ is
\begin{equation}
\phi'(u,\lambda) = \frac{1}{i}\frac{\di}{\di u}\log \frac{\sinh(u-\I\lambda)}
{\sinh(u+\I\lambda)} = \frac{2\sin 2\lambda}{\cosh 2u - \cos 2\lambda}.
\end{equation}
After taking logarithms, the Bethe ansatz equations (\ref{eqn:honeybae})
(and their
analogues for the other boundary conditions) can be rewritten as
\begin{eqnarray}
\lefteqn{ N \left[ \phi(u_k,\smallfrac{1}{2}\lambda)+
 \phi(u_k,\smallfrac{3}{2}\lambda)\right] + \phi_{\rm bound}(u_k,\lambda)
 = 2 \pi I_k + }
\hspace{10pt}\nonumber\\
&& \sum_{j\neq k}^M \left[ \phi(u_k+u_j,2\lambda)+\phi(u_k-u_j,2\lambda)-
\phi(u_k+u_j,\lambda)-\phi(u_k-u_j,\lambda)\right].
\label{eqn:logbae}
\end{eqnarray}
Here the set of integers $I_k$ for $k=1,\ldots,M$ uniquely
specifies an eigenstate of the transfer matrix. The term
$\phi_{\rm bound}(u,\lambda)$ depends on the boundary conditions chosen and
is explicitly given by
\begin{equation}
\phi_{\rm bound}(u,\lambda) = \left\{ \begin{array}{cc}
  2 \phi(u,\smallfrac{1}{2}\lambda) & {\rm O-O}\\
  \phi(2u,\lambda) & {\rm O-Sp}\\
  2 \phi(u+\I\smallfrac{\pi}{2},\smallfrac{1}{2}\lambda) & {\rm Sp-Sp}
\end{array}.\right.
\label{eqn:bound}
\end{equation}

We determined the values of $I_k$ for the eigenvalues $\Lambda_{\ell}$ of
interest numerically, i.e. by substituting the values of the Bethe ansatz
roots $u_j$, determined from the numerical solution of the Bethe ansatz
equations (\ref{eqn:honeybae}), into the logarithmic form (\ref{eqn:logbae}).
We expect that they can also be obtained by a careful consideration of the
branch cuts involved in passing to the logarithmic form (see, e.g.\
\cite{Baxter86}). The following identification is found for those
eigenstates corresponding to real Bethe ansatz roots:
\begin{eqnarray}
\Lambda_{\ell}^{\rm O-O} &\leftrightarrow& I_k=k,\;\;k=1,\ldots,M=N-\ell,
\nonumber\\
\Lambda_{\ell}^{\rm O-Sp} &\leftrightarrow& I_k=k,\;\;k=1,\ldots,M=N-\ell,
\nonumber\\
\Lambda_{\ell}^{\rm Sp-Sp} &\leftrightarrow& I_k=k+1,\;\;k=1,\ldots,M=N-\ell
\hspace{10pt} (\ell\neq 0).
\label{eqn:iden}
\end{eqnarray}
For the eigenvalue $\Lambda^{\rm Sp-Sp}_0$ we find that $I_j=j$ for the real
roots $j=2,\ldots,N$.

Define the ``counting function'' $z_N(u)$ by
\begin{equation}
2\pi z_N(u)= \phi(u,\smallfrac{1}{2}\lambda) +
\phi(u,\smallfrac{3}{2}\lambda) -
\frac{1}{N}\sum_{j=-M}^{M} \left[ \phi(u-u_j,2\lambda)-
\phi(u-u_j,\lambda)\right] + \frac{1}{N}\Phi(u),
\label{eqn:count}
\end{equation}
where
\begin{equation}
\Phi(u) = \phi_{\rm bound}(u,\lambda) + \phi(u,2\lambda) - \phi(u,\lambda) +
\phi(2u,2\lambda)-\phi(2u,\lambda)
\label{eqn:phi}
\end{equation}
contains the boundary contribution.
For convenience, we have symmetrized $z_N(u)$ through the identification
$u_{-k} \equiv -u_k$ for $k>0$ and $u_0 \equiv 0$.
Using the Bethe ansatz equations (\ref{eqn:logbae}) it can be verified that
for Bethe ansatz roots $u_k$ we have
\begin{equation}
z_N(u_k) = \frac{I_k}{N},
\end{equation}
a property for which $z_N(u)$ is designed.
Therefore, with the values of $I_k$ given in (\ref{eqn:iden}),
\begin{equation}
\rho_N(u)\equiv \frac{\di z_N(u)}{\di u}
\end{equation}
becomes a root density for large $N$. Using this
we can approximate for large $N$ (and therefore also $M$)
\begin{equation}
\frac{1}{N}\sum_{k=-M}^M f(u_k) \approx \int_{-\infty}^{\infty} \di v
\rho_{\infty}
(v) f(v).
\end{equation}

\subsection{Bulk free energy}
In the limit $N \rightarrow \infty$ Eq.\ (\ref{eqn:count}) gives rise
to the linear integral equation
\begin{equation}
2 \pi \rho_{\infty}(u) = \phi'(u,\smallfrac{1}{2}\lambda) +
\phi'(u,\smallfrac{3}{2}\lambda)-
\int_{-\infty}^{\infty} \di v \rho_{\infty}(v)
\left[ \phi'(u-v,2\lambda) - \phi'(u-v,\lambda) \right]
\label{eqn:inteq1}
\end{equation}
for the root density $\rho_{\infty}(u)$.
This equation is readily solved by Fourier transforms. We use henceforth
the notation
\begin{eqnarray}
\widehat{a}(x) & = &\frac{1}{2\pi}\int_{-\infty}^{\infty} \di u e^{\I x u}
 a(u),\nonumber\\
a(u) & = & \int_{-\infty}^{\infty} \di x e^{-\I x u} \widehat{a}(x).
\end{eqnarray}
Using the result
\begin{equation}
\widehat{\phi'}(x,\lambda) = \frac{\sinh[(\smallfrac{\pi}{2}-\lambda)x]}
{\sinh \smallfrac{1}{2}\pi x}
\hspace{20pt} \left(0<\lambda<\smallfrac{\pi}{2}\right),
\label{eqn:ftphi}
\end{equation}
which can be established by contour integration, we find that the root
density is given by
\begin{equation}
2 \pi \widehat{\rho}_{\infty}(x)
= \frac{2 \cosh\smallfrac{1}{2}\lambda x}{2 \cosh \lambda x -1}.
\label{eqn:rden}
\end{equation}
The eigenvalue expression (\ref{eqn:honeyev}) can be rewritten in the
symmetrized form
\begin{equation}
\Lambda_{\rm H} = \frac{\sinh(\smallfrac{1}{2}\I \lambda)}
{\sinh(\smallfrac{3}{2}\I \lambda)}\prod_{j=-M}^{M} \frac{\sinh(u_j-
\smallfrac{3}{2}\I\lambda)}{\sinh(u_j-\smallfrac{1}{2}\I\lambda)}.
\end{equation}
Let us denote for convenience
\begin{equation}
\psi(u) = \log \frac{\sinh(u-\smallfrac{3}{2}\I\lambda)}{\sinh(u-
  \smallfrac{1}{2}\I\lambda)}.
\end{equation}
Then the free energy per site, defined as
$f_N \equiv - \log \Lambda_{\rm H}/N$, becomes
in the large $N$ limit
\begin{eqnarray}
f_{\infty} & = &- \int_{-\infty}^{\infty} \di v \rho_{\infty}(v)\psi(v)
\nonumber\\
&=& -\int_{-\infty}^{\infty} \di x \frac{\sinh[(\smallfrac{\pi}{2}-\lambda)x]
   \sinh(\lambda x)}
{x \sinh\smallfrac{1}{2}\pi x \left( 2 \cosh\lambda x -1\right)}.
\end{eqnarray}
The last equality is obtained using (\ref{eqn:rden}) and the Fourier transform
\begin{equation}
\widehat{\psi}(x) = \frac{e^{(\pi-2\lambda)x/2}\sinh\smallfrac{1}{2}\lambda x}
{x \sinh\smallfrac{1}{2}\pi x}
\hspace{20pt} \left(0<\lambda<\smallfrac{2\pi}{3}\right),
\end{equation}
which is also readily established by a contour integration involving its
derivative.  Not surprisingly, the boundary terms have not played a
part up to this point. Indeed, the bulk free energy is the same if one starts
from periodic boundary conditions \cite{Baxter86}. In the next few sections we
deal with the finite-size corrections in which the boundary terms are
important.

\subsection{Finite-size corrections}
Let us define now the functions
\begin{eqnarray}
P_N(u) &=& \rho_N(u) - \rho_{\infty}(u),\\
S_N(u) &=& \frac{1}{N}\sum_{k=-M}^{M} \delta(u-u_k) - \rho_N(u),\\
K(u) & = & \phi'(u,2\lambda)-\phi'(u,\lambda).
\end{eqnarray}
On differentiating (\ref{eqn:count}) and using (\ref{eqn:inteq1})  we
find that $P_N(u)$ satisfies the integral equation
\begin{eqnarray}
\lefteqn{P_N(u) + \frac{1}{2\pi} \int_{-\infty}^{\infty} \di v P_N(v)
K(u-v)}
\hspace{70pt} \nonumber\\
&=& -\frac{1}{2\pi} \int_{-\infty}^{\infty} \di v
K(u-v) S_N(v) + \frac{1}{2\pi N}
\Phi'(u).
\label{eqn:pn1}
\end{eqnarray}
Taking the Fourier transform of both sides, gathering terms and then taking
the inverse Fourier transform results in the following result for $P_N(u)$:
\begin{eqnarray}
\lefteqn{
P_N(u)= -\frac{1}{(2\pi)^2}\int_{-\infty}^{\infty} \di w
\int_{-\infty}^{\infty} \di v K(v-w) S_N(w) K_1(u-v) }
\hspace{70pt}\nonumber\\
& + &
\frac{1}{(2\pi)^2 N}\int_{-\infty}^{\infty} \di v \Phi'(v)K_1(u-v),
\end{eqnarray}
where $K_1(u)$ is defined such that $\widehat{K}_1(x)=\left(1+\widehat{K}(x)
\right)^{-1}$. This can be given a different representation in terms of
a function $K_2(u)$ defined such that $\widehat{K}_2(x)=-\widehat{K}(x)
\widehat{K}_1(x)$:
\begin{equation}
P_N(u)= \frac{1}{2\pi}\int_{-\infty}^{\infty} \di w
S_N(w) K_2(u-w)
+\frac{1}{(2\pi)^2 N}\int_{-\infty}^{\infty} \di v \Phi'(v)K_1(u-v).
\label{eqn:pn}
\end{equation}
The Fourier transforms $\widehat{K}(x)$,$\widehat{K}_1(x)$, and thus
$\widehat{K}_2(x)$ follow from (\ref{eqn:ftphi}):
\begin{eqnarray}
\widehat{K}(x) &=& - \frac{2\cosh\left[\smallfrac{1}{2}
(\pi-3\lambda)x\right] \sinh\smallfrac{1}{2}
x\lambda}
  {\sinh\smallfrac{1}{2}\pi x},\nonumber\\
\widehat{K}_1(x) &=& \frac{\sinh\smallfrac{1}{2}\pi x}{(2\cosh\lambda x-1)
\sinh[(\smallfrac{\pi}{2}-\lambda)x]}.
\label{eqn:kk1}
\end{eqnarray}

Analogous to Eq.\ (\ref{eqn:pn1}) we find the following expression for the
finite-size corrections to the free energy from the definitions of $f_N$ and
$f_{\infty}$
\begin{eqnarray}
f_N-f_{\infty} & = & -\frac{1}{2N} \log \left(
\frac{1-\cos \lambda}{1-\cos 3\lambda}\right)
  - \int_{-\infty}^{\infty} \di v P_N(v) \psi(v) \nonumber\\
 &-& \int_{-\infty}^{\infty} \di v S_N(v) \psi(v).
\end{eqnarray}
Substituting in the result (\ref{eqn:pn}) for $P_N(u)$, we find that
\begin{eqnarray}
f_N-f_{\infty} & = & -\frac{1}{2N} \log \left(
\frac{1-\cos \lambda}{1-\cos 3\lambda}\right)
  - \frac{1}{4\pi^2 N} \int_{-\infty}^{\infty} \di u
\int_{-\infty}^{\infty} \di v \Phi'(v) \psi(v) K_1(u-v) \nonumber\\
   &-& \int_{-\infty}^{\infty} \di v S_N(v) e_2(v),
\label{eqn:fninf}
\end{eqnarray}
where $e_2(u)$ is defined such that
\begin{equation}
\widehat{e}_2(x)= \frac{\sinh\smallfrac{1}{2}\lambda x}
{x(2 \cosh\lambda x-1)}.
\end{equation}
In fact, by contour integration, this can be inverted to give ($\lambda>0$)
\begin{equation}
e_2(u) = \frac{1}{2}\log \left( \frac{2 \cosh\frac{\pi u}{3\lambda}+ \sqrt{3}}
  {2 \cosh\frac{\pi u}{3\lambda}-\sqrt{3}}\right).
\label{eqn:e2}
\end{equation}
Eqs.\ (\ref{eqn:pn}) and (\ref{eqn:fninf}) are key equations\footnote{There
are misprints in the corresponding equations given in
\cite{Batchelor93} for O-O boundaries.}
in what follows
and lead directly to the various terms in the free energy relation
(\ref{eqn:confinv}). Observe that in deriving these equations we have not had
to specify in which sector of the transfer matrix we work. If we choose to be
in sector $\ell$ then $f_N$ is to be interpreted as the lowest free energy
in that sector. The scaling behaviour (\ref{eqn:confinv}) in sector $0$ leads
to the central charge while that in sector $\ell$ leads to the relevant
scaling dimensions.

\subsection{Surface free energies}
The $O(1/N)$ terms in Eq.\ (\ref{eqn:fninf}) give rise to the
surface energy $f_{\rm S}$. Using the Fourier transform (\ref{eqn:ftphi}) we
can evaluate $\widehat{\Phi'}(x)$ for each type of boundary condition under
consideration using the definitions (\ref{eqn:phi}) and (\ref{eqn:bound}). For
O-O boundaries we recover a result of \cite{Batchelor93}
\begin{eqnarray}
\lefteqn{
f_{\rm S}^{\rm O-O} = -\frac{1}{2}\log \left(
\frac{1-\cos \lambda}{1-\cos 3\lambda}\right)
}\hspace{20pt}\nonumber\\
&-& 4 \int_{-\infty}^{\infty} \di x \frac{\sinh\smallfrac{\lambda x}{2}
  \cosh\smallfrac{\lambda x}{4} \cosh\smallfrac{(\pi-2\lambda)x}{4}
  \sinh\smallfrac{(\pi-3\lambda)x}{4}
 \left(2 \cosh\smallfrac{\lambda x}{2}-1\right)}
{x \sinh\smallfrac{\pi x}{2} \left(2 \cosh\lambda x-1\right)}.
\label{eqn:fsoo}
\end{eqnarray}
For O-Sp boundaries we find
\begin{eqnarray}
\lefteqn{
f_{\rm S}^{\rm O-Sp}= -\frac{1}{2}\log \left(
\frac{1-\cos \lambda}{1-\cos 3\lambda}\right)
}\hspace{20pt}\nonumber\\
&-& 2\int_{-\infty}^{\infty} \di x \frac{\sinh\smallfrac{\lambda x}{2}
  \cosh\smallfrac{(\pi-2\lambda)x}{4} \sinh\smallfrac{(\pi-6\lambda)x}{4}}
{x \sinh\smallfrac{\pi x}{2} \left(2 \cosh\lambda x-1\right)}.
\label{eqn:fsos}
\end{eqnarray}
For Sp-Sp boundaries we use the Fourier transform
\begin{equation}
\frac{1}{2\pi}\int_{-\infty}^{\infty} \di u e^{\I x u} \phi'(u+\I
\smallfrac{\pi}{2},\smallfrac{\lambda}{2}) = -\frac{\sinh\smallfrac{1}{2}
\lambda x}
{\sinh\smallfrac{1}{2}\pi x} \hspace{20pt}
\left(0<\lambda<\smallfrac{\pi}{2}\right),
\end{equation}
which can also be derived using contour integration, to obtain
\begin{eqnarray}
\lefteqn{
f_{\rm S}^{\rm Sp-Sp}= -\frac{1}{2}\log \left(
\frac{1-\cos \lambda}{1-\cos 3\lambda}\right)
}\hspace{20pt}\nonumber\\
&+& 4\int_{-\infty}^{\infty} \di x \frac{\sinh\smallfrac{\lambda x}{2}
\sinh\smallfrac{\lambda x}{4} \cosh\smallfrac{(\pi-3\lambda)x}{4}
\cosh\smallfrac{(\pi-2\lambda)x}{4}\left(1+2\cosh\smallfrac{\lambda x}{2}
\right)}
{x \sinh\smallfrac{\pi x}{2} \left(2 \cosh\lambda x-1\right)}.
\label{eqn:fsss}
\end{eqnarray}
Non-trivial numerical checks can be performed at this stage. {F}rom
numerical transfer matrix calculations we find that the largest eigenvalue
$\Lambda_0$ of the transfer matrix at $\lambda=\pi/8$ (i.e.\ the dilute
polymer point $n=0$) is given by
\begin{equation}
\Lambda_0 = \left(2+\sqrt{2}\right)^N \left\{
\begin{array}{cc}
1 & {\rm O-O}\\
\left(1+\sqrt{2}\right)^{-1} & {\rm O-Sp}\\
\left(1+\sqrt{2}\right)^{-2} & {\rm Sp-Sp}
\end{array}
\right.,
\end{equation}
without further corrections in $1/N$.
Therefore the surface free energies at $\lambda=\pi/8$ are given
respectively by $0$, $\log(1+\sqrt{2})$ and $2\log(1+\sqrt{2})$. This is
confirmed by numerical integration of Eqs.\ (\ref{eqn:fsoo}),
(\ref{eqn:fsos}) and (\ref{eqn:fsss}).

It is not hard to show that the surface energies for general $\lambda$
are related by
\begin{equation}
f_{\rm S}^{\rm O-Sp}= \smallfrac{1}{2}\left(f_{\rm S}^{\rm O-O}+
 f_{\rm S}^{\rm Sp-Sp}\right),
\end{equation}
which is, of course, a consequence of the relation (\ref{eqn:relth})
amongst the corresponding boundary functions. A comparison with
Eq.\ (\ref{eqn:confinv}) suggests
sharing out the surface free energies in the following way:
\begin{eqnarray}
f_{\rm S}^{({\rm O})} &=& \smallfrac{1}{2} f_{\rm S}^{\rm O-O}\nonumber\\
f_{\rm S}^{({\rm S})} &=& \smallfrac{1}{2} f_{\rm S}^{\rm Sp-Sp}.
\end{eqnarray}

\subsection{Wiener-Hopf integral equation}
We move on now to the next leading order in $1/N$ in the finite-size
correction $f_N-f_{\infty}$.
The Euler-Maclaurin formula gives rise to
\begin{eqnarray}
\int_{-\infty}^{\infty} \di u g(u) S_N(u) & \approx & -\left(
  \int_{-\infty}^{-\Lambda} \di u + \int_{\Lambda}^{\infty} \di u \right)
  g(u) \rho_N(u) + \nonumber\\
 & & \frac{1}{2N}\left( g(\Lambda) + g(-\Lambda) \right) +
  \frac{1}{12 N^2 \rho_N(\Lambda)} \left( g'(\Lambda) + g'(-\Lambda) \right),
\label{eqn:euler}
\end{eqnarray}
for arbitrary $g(u)$ analytic in $[-\Lambda,\Lambda]$ where $\Lambda$ is the
largest root. In other words, $\Lambda$ is such that
$z_N(\Lambda)=I_{\rm max}/N$ where, on using (\ref{eqn:iden}),
$I_{\rm max}=N-\ell$ in sector $\ell$ for O-O and O-Sp boundaries,
and $I_{\rm max}=N-\ell+1$ in sector $\ell \neq 0$ for Sp-Sp boundaries.
The condition on $\Lambda$ leads to the sum rule\footnote{This
corrects a misprint in the sum rule for O-O boundaries
given in \cite{Batchelor93}}
\begin{eqnarray}
\int_{\Lambda}^{\infty} \di u \rho_N(u) & = & z_N(\infty) - z_N(\Lambda)
\nonumber\\
&=& 1+ \frac{2\lambda}{\pi}\left(\frac{M}{N}-1\right) - \frac{\lambda}{\pi N}
+ \frac{\phi_{\rm bound}(\infty,\lambda)}{2\pi N} - \frac{I_{\rm max}}{N},
\label{eqn:sumrule}
\end{eqnarray}
where the last equality follows from the definition (\ref{eqn:count}) and the
result $\phi(\infty,\lambda)=\pi-2\lambda$.
Applying (\ref{eqn:euler}) to the first integral of Eq.\ (\ref{eqn:pn}) we
obtain
\begin{eqnarray}
\rho_N(u) - \rho_{\infty}(u) & \approx & \frac{1}{4\pi^2 N}
\int_{-\infty}^{\infty} \di v K_1(u-v) \Phi'(v) - \frac{1}{2\pi}
\int_{\Lambda}^{\infty} \di v \rho_N(v) K_2(u-v) + \nonumber\\
& & \frac{1}{4\pi N} K_2(u-\Lambda)- \frac{1}{24 \pi N^2 \rho_N(\Lambda)}
K_2'(u-\Lambda),
\label{eqn:prewh}
\end{eqnarray}
which is valid for $u \geq \Lambda$. Under this condition, certain terms
implied by (\ref{eqn:euler}) are negligible because $K_2(u)$ decreases
exponentially with
increasing $u$ and have been dropped. Applying (\ref{eqn:euler})
to the last term of Eq.\ (\ref{eqn:fninf}) and using the fact that $e_2(u)$
is an even function we obtain also
\begin{equation}
-\int_{-\infty}^{\infty} \di v S_N(v) e_2(v)  \approx
2 \int_{\Lambda}^{\infty} \di u \rho_N(u) e_2(u) -
 \frac{1}{N} e_2(\Lambda) - \frac{e_2'(\Lambda)}{6N^2 \rho_N(\Lambda)}.
\label{eqn:fseuler}
\end{equation}
This term describes the $O(1/N^2)$ correction to the free energy and in the
thermodynamic limit is governed by the behaviour of $\rho_N(\Lambda)$ for
large $\Lambda$. This in turn can be found by solving Eq.\ (\ref{eqn:prewh})
subject to the sum rule (\ref{eqn:sumrule}), as we now demonstrate.

Define the functions
\begin{eqnarray}
\chi(u) &=& \rho_N(u+\Lambda),\nonumber\\
f(u) &=& \rho_{\infty}(u+\Lambda),\nonumber\\
\epsilon(u) &=& \frac{1}{4\pi^2N} \int_{-\infty}^{\infty} \di v
K_1(u+\Lambda-v)
\Phi'(v).
\label{eqn:defchi}
\end{eqnarray}
The integral equation (\ref{eqn:prewh}) can then be rewritten as
\begin{equation}
\chi(u) \approx f(u) + \epsilon(u) - \frac{1}{2\pi} \int_0^{\infty} \di v
\chi(v) K_2(u-v) + \frac{1}{4\pi N} K_2(u) - \frac{K_2'(u)}{24\pi N^2
\rho_N(\Lambda)}.
\label{eqn:wh}
\end{equation}
This can be recognised as a Wiener-Hopf type integral equation for
$\chi(u)$ which can be solved by standard means.

First split $\widehat{\chi}(x)=\widehat{\chi}_+(x)+\widehat{\chi}_-(x)$ into
components analytic in the upper and lower half-planes $\Pi_{\pm}$, with
\begin{equation}
\widehat{\chi}_+(x) = \frac{1}{2\pi}\int_0^{\infty} \di u
e^{\I x u} \chi(u).
\label{eqn:chiplus}
\end{equation}
Then the Fourier transform of the integral equation (\ref{eqn:wh}) can be
written as
\begin{equation}
\widehat{\chi}(x) \approx \widehat{f}(x) +\widehat{\epsilon}(x) -
\widehat{\chi}_+(x) \widehat{K}_2(x) + \frac{1}{2\pi}\widehat{K}_2(x) C(x),
\label{eqn:wh1}
\end{equation}
where we define
\begin{equation}
C(x) \equiv \frac{1}{2N} + \frac{\I x}{12N^2 \rho_N(\Lambda)}.
\label{eqn:cx}
\end{equation}
Let $F(x)\equiv \widehat{f}(x) + \widehat{\epsilon}(x)$ be split into
components analogous to $\widehat{\chi}(x)$. Then (\ref{eqn:wh1}) can be
rewritten as
\begin{equation}
\widehat{\chi}_-(x) + \left[1+\widehat{K}_2(x)\right] \left[
\widehat{\chi}_+(x) - \smallfrac{1}{2\pi}C(x) \right]
\approx F_+(x) + F_-(x) - \smallfrac{1}{2\pi}C(x).
\label{eqn:wh2}
\end{equation}
Now, if we can factorize
\begin{equation}
\left[1+\widehat{K}_2(x)\right]^{-1} = G_+(x) G_-(x),
\label{eqn:fact}
\end{equation}
where $G_+(x)$ ($G_-(x)$) is analytic in $\Pi_+$ (respectively, $\Pi_-$),
and $G_-(x) F_+(x)$ can be split into $+$ and $-$ components
\begin{equation}
G_-(x)F_+(x)= Q_+(x) + Q_-(x),
\label{eqn:dec2}
\end{equation}
then Eq.\ (\ref{eqn:wh2}) can be written in the form
\begin{eqnarray}
\frac{\widehat{\chi}_+(x)-\smallfrac{1}{2\pi}C(x)}{G_+(x)} - Q_+(x) & = &
Q_-(x) - G_-(x) \left[ \widehat{\chi}_-(x)+\smallfrac{1}{2\pi}
C(x)-F_-(x)\right]\nonumber\\
& \equiv & P(x).
\label{eqn:reg}
\end{eqnarray}
Since the first two terms are each analytic in $\Pi_+$ and
$\Pi_-$ respectively, $P(x)$ must be a regular function and can be
determined, for instance, from the large $x$ behaviour of the first term.

Let us now confirm the decompositions (\ref{eqn:fact}) and (\ref{eqn:dec2}).
Firstly, we find from (\ref{eqn:kk1}) that
\begin{equation}
\left[1+\widehat{K}_2(x)\right]^{-1}
= \frac{\left(2\cosh\lambda x -1\right)
\sinh[(\pi-2\lambda)x/2]}{\sinh(\pi x/2)}.
\end{equation}
It indeed factorizes as (\ref{eqn:fact}) with \cite{Batchelor88,Suzuki88}
\begin{eqnarray}
G_+(x) & = & \frac{2\sqrt{\pi(\pi-2\lambda)}\Gamma\left(1-\I\smallfrac{x}{2}
\right) e^{h(x)}}
{\Gamma\left(1-\smallfrac{\I x}{2\pi}(\pi-2\lambda)\right)
\Gamma\left(\smallfrac{1}{6}-\smallfrac{\I\lambda}{2\pi}x\right)
\Gamma\left(\smallfrac{5}{6}-\smallfrac{\I\lambda}{2\pi}x\right)}
\nonumber\\
&=& G_-(-x),
\label{eqn:rh}
\end{eqnarray}
where
\begin{equation}
h(x)= \frac{\I x}{2} \left[ \log \frac{\pi}{\lambda} -
\smallfrac{(\pi-2\lambda)}{\pi} \log \frac{\pi-2\lambda}{\lambda}\right].
\end{equation}
As $|x| \rightarrow \infty$ in $\Pi_+$ we find, on using Stirling's
formula, that
\begin{equation}
G_+(x) \sim 1+\frac{g_1}{x} + \frac{g_2}{x^2}
+ O(x^{-3}),
\label{eqn:gpx}
\end{equation}
where the coefficients are given by
\begin{eqnarray}
g_1 &=& \frac{\I}{3}\left[ \frac{1}{2} - \frac{\pi}{2(\pi-2\lambda)}
-\frac{\pi}{6\lambda} \right],\nonumber\\
g_2 &=& \smallfrac{1}{2} g_1^2.
\end{eqnarray}
Next, from the definition (\ref{eqn:defchi}) and the result (\ref{eqn:rden})
we have
\begin{equation}
\widehat{f}(x) = \frac{e^{-\I x \Lambda}} {2\pi}
\frac{2 \cosh(\lambda x/2)}{(2\cosh \lambda x -1 )},
\end{equation}
whose leading pole in $\Pi_-$ is at $ x=-\I \pi/3\lambda$. Keeping only
this leading pole term we obtain
\begin{equation}
\widehat{f}_+(x) \approx  \frac{e^{-\I x \Lambda}} {2\pi}
\frac{2\sqrt{3} \cosh(\lambda x/2)}{(\pi-3\I\lambda x)} \approx
F_+(x).
\end{equation}
Here we have ignored the contribution to $F(x)$ from $\epsilon(x)$
(cf.\ Eq.\ (\ref{eqn:defchi})), which
contributes to the finite-size corrections at a higher order in $1/N$.
One can now see that $G_-(x)F_+(x)$ has only one pole in $\Pi_-$, at
$x=-\I \pi/3\lambda$ and use this fact to obtain
\begin{equation}
Q_+(x)= \frac{3 G_+\left(\frac{\I\pi}{3\lambda}\right)e^{-\frac{\pi \Lambda}
{3\lambda}}}{2\pi(\pi-3\I\lambda x)}.
\label{eqn:qpx}
\end{equation}
In particular, $Q_+(x) \rightarrow 0$ as $|x|\rightarrow \infty$ in $\Pi_+$.
In the same limit, $\widehat{\chi}_+(x) \rightarrow 0$ from the requirement
that $\chi_+(u)$ is integrable at the origin. Therefore from Eqs.\
(\ref{eqn:reg}), (\ref{eqn:cx}) and (\ref{eqn:gpx}) we obtain
\begin{equation}
P(x)= -\frac{1}{2\pi}\left[ \frac{1}{2N} + \frac{\I(x-g_1)}{12N^2
\rho_N(\Lambda)}\right].
\label{eqn:px}
\end{equation}

Knowing $C(x)$, $G_+(x)$, $Q_+(x)$ and $P(x)$, we are then in a position to
obtain $\widehat{\chi}_+(x)$ using Eq.\ (\ref{eqn:reg}).
We now note from the definitions (\ref{eqn:defchi}) and (\ref{eqn:chiplus})
that $\widehat{\chi}_+(x)=\smallfrac{1}{2\pi} \int_{0}^{\infty} \di u e^{\I x
u}
\rho_N(u+\Lambda)$. Therefore from Eq.\ (\ref{eqn:reg}) we obtain
\begin{equation}
\frac{e^{-\I x \Lambda}}{2\pi} \int_{\Lambda}^{\infty}
\di u e^{\I x u} \rho_N(u)
 = G_+(x) \left[ P(x) + Q_+(x) \right] + \frac{1}{2\pi} C(x).
\label{eqn:intrho}
\end{equation}
{F}rom this equation, two key results (\ref{eqn:quad}) and
(\ref{eqn:alpha2}) will be obtained.
We note at this point that since we have dropped $\widehat{
\epsilon}(x)$ -- which contains the contribution from $\Phi(u)$ and therefore
$\phi_{\rm bound}(u)$ -- from $F(x)$, Eq.\ (\ref{eqn:intrho}) does not
depend on the boundary conditions.

Taking the inverse Fourier transform of $\widehat{\chi}_+(x)$, we obtain
$\rho_N(\Lambda)$ as a contour integral, which can be written as a residue
at infinity:
\begin{equation}
\chi_+(0)=\frac{1}{2}\rho_N(\Lambda)= -\I\pi \lim_{x\rightarrow\infty} \left[
x G_+(x)\left[ P(x)+Q_+(x) \right] + \frac{x}{2\pi} C(x) \right].
\label{eqn:rhon}
\end{equation}
We now subsititute Eqs.\ (\ref{eqn:cx}), (\ref{eqn:px}), (\ref{eqn:gpx}) and
(\ref{eqn:qpx}) into (\ref{eqn:rhon}) and evaluate the limit to find
\begin{eqnarray}
N \rho_N(\Lambda) &=& \frac{\I g_1}{2} + \frac{g_1^2-g_2}{12N \rho_N(\Lambda)}
+ \frac{2\pi^2N}{3\lambda}\left[Q_+(0)+P(0)\right] + \nonumber\\
& & \frac{\pi}{3\lambda}
 \left[\frac{1}{2}-\frac{\I g_1}{12N \rho_N(\Lambda)}\right].
\label{eqn:nrhon}
\end{eqnarray}
In arriving at the final result we have eliminated $G_+(\I\pi/3\lambda)$ in
favour of $Q_+(0)$ using (\ref{eqn:qpx}).
Define
\begin{equation}
\rho\equiv N\rho_N(\Lambda), \hspace{40pt}
\alpha\equiv N\left[Q_+(0)+P(0) \right].
\end{equation}
 Then Eq.\ (\ref{eqn:nrhon}) can be succintly written as a
quadratic equation for $\rho$
\begin{equation}
\rho^2 = \rho \left( \frac{\I g_1}{2} + \frac{2\pi^2}{3\lambda}\alpha +
\frac{\pi}{6\lambda}\right) + \frac{g_1^2-g_2}{12}-\frac{\I\pi g_1}{36\lambda}.
\label{eqn:quad}
\end{equation}
We will see in the next section that
the $O(N^{-2})$ correction to the free energy can
be expressed in terms of $\rho$, $\alpha$ and $g_1$. We will also see that
on using the quadratic equation (\ref{eqn:quad}) this expression becomes
just a function of $\alpha^2$.

Let us now relate the evaluation of $\alpha$ to the sum rule
(\ref{eqn:sumrule}). On setting $x=0$ in  Eq.\ (\ref{eqn:intrho}) we obtain
\begin{equation}
\int_{\Lambda}^{\infty} \di u \rho_N(u) = 2\pi G_+(0) \left[ P(0)+Q_+(0)
\right] + C(0).
\label{eqn:compsr}
\end{equation}
{}From (\ref{eqn:rh}) we have
\begin{equation}
G_+(0)=\frac{2\sqrt{\pi(\pi-2\lambda)}}{\Gamma\left(\smallfrac{1}{6}\right)
\Gamma\left(\smallfrac{5}{6}\right)}=\frac{\sqrt{\pi(\pi-2\lambda)}}{\pi},
\end{equation}
and from (\ref{eqn:cx}) we can evaluate $C(0)$. A comparison of
(\ref{eqn:compsr}) with the sum rule (\ref{eqn:sumrule}) then leads to the
result
\begin{eqnarray}
\lefteqn{4 \pi^2 \alpha^2 = }\nonumber\\
& & \left(1-\frac{2\lambda}{\pi}\right)^{-1}
\left[
N - \frac{1}{2}
+ \frac{2\lambda}{\pi}\left(M-N\right) - \frac{\lambda}{\pi }
+ \frac{\phi_{\rm bound}(\infty,\lambda)}{2\pi } - I_{\rm max}
\right]^2,
\label{eqn:alpha2}
\end{eqnarray}
where the boundary contribution has finally reappeared.

\subsection{Central charge and scaling dimensions}
We are now ready to return to the finite-size correction to the free energy
(\ref{eqn:fseuler}). From Eq.\ (\ref{eqn:e2}) we have
\begin{equation}
e_2(u) \sim \sqrt{3} e^{-\frac{\pi u}{3\lambda}}
\end{equation}
for large $u$. Substitution into Eq.\ (\ref{eqn:fseuler}) yields
\begin{eqnarray}
\lefteqn{- \int_{-\infty}^{\infty} \di v S_N(v) e_2(v) \approx}\nonumber\\
& &
2\sqrt{3}\int_{\Lambda}^{\infty} \di u e^{\I u \left( \frac{\I\pi}{3\lambda}
\right)} \rho_N(u) - \left[ \frac{\sqrt{3}}{N}
-\frac{\sqrt{3}\pi}{18\lambda N^2 \rho_N(\Lambda)}\right]
 e^{-\frac{\pi\Lambda}{3\lambda}}.
\end{eqnarray}
Using the equation (\ref{eqn:intrho}) with $x=\I\pi/3\lambda$ and the
expression (\ref{eqn:cx}) for $C(x)$ we end up with
\begin{equation}
- \int_{-\infty}^{\infty} \di v S_N(v) e_2(v) \approx
4\sqrt{3}\pi e^{-\frac{\pi\Lambda}{3\lambda}} G_+
\left(\smallfrac{\I\pi}{3\lambda} \right)
\left[ P \left(\smallfrac{\I\pi}{3\lambda}\right) +
Q_+  \left(\smallfrac{\I\pi}{3\lambda}\right) \right].
\end{equation}
Both $G_+\left(\smallfrac{\I\pi}{3\lambda} \right)$ and
$Q_+ \left(\smallfrac{\I\pi}{3\lambda}\right)$ can be eliminated in favour
of $Q_+(0)$ using (\ref{eqn:qpx}). This in turn can be expressed in terms
of $\alpha$ and $\rho$. We have thus
\begin{eqnarray}
- \int_{-\infty}^{\infty} \di v S_N(v) e_2(v) & \approx &
\frac{4\sqrt{3}\pi}{N^2}\left[ \frac{2\pi^2}{3}\alpha +
\frac{\pi}{3}\left(\frac{1}{2}-\frac{\I g_1}{12\rho}\right)\right]\nonumber\\
& \times &
\left( \frac{\alpha}{2}-\frac{1}{8\pi} +\frac{\I g_1}{48\pi \rho}+
\frac{1}{72\lambda \rho}\right).
\end{eqnarray}
Expanding out the terms and using the quadratic equation (\ref{eqn:quad})
for $\rho$ we find eventually that
\begin{equation}
- \int_{-\infty}^{\infty} \di v S_N(v) e_2(v) \approx
-\frac{\pi}{24 N^2}\left(\frac{2}{\sqrt{3}}\right)
\left( 1- 48\pi^2 \alpha^2 \right).
\label{eqn:on2}
\end{equation}

Now from Eq.\ (\ref{eqn:alpha2}) we have the result
\begin{equation}
4\pi^2\alpha^2 = \left(1-\smallfrac{2\lambda}{\pi}\right)^{-1}
\left\{\begin{array}{ccc}
\left[ \ell\left(1-\smallfrac{2\lambda}{\pi}\right)-\smallfrac{2\lambda}{\pi}+
\smallfrac{1}{2}\right]^2 & {\rm O-O} &\\[5pt]
\left[ \ell\left(1-\smallfrac{2\lambda}{\pi}\right)-\smallfrac{2\lambda}{\pi}
\right]^2 & {\rm O-Sp}&\\[5pt]
\left[ \ell\left(1-\smallfrac{2\lambda}{\pi}\right)-\smallfrac{2\lambda}{\pi}-
\smallfrac{1}{2}\right]^2 & {\rm Sp-Sp} & (\ell \neq 0)
\end{array}\right.
\label{eqn:alpha3}
\end{equation}
for the various eigenstates of interest, except for the ground state with
Sp-Sp boundary conditions. Let us first consider the ground states ($\ell=0$)
for the O-O and O-Sp cases. Substituting the result for $\alpha^2$ in Eq.\
(\ref{eqn:alpha3}) into Eq.\ (\ref{eqn:on2}) and comparing with the
result (\ref{eqn:confinv}) leads to the identification
$\zeta=2/\sqrt{3}$ for the scale factor\footnote{This
corrects misprints in \cite{Batchelor93,Batchelor95a}},
together with the central charges
\begin{eqnarray}
c^{\rm O-O} &=& 1- \frac{3(\pi-4\lambda)^2}{\pi(\pi-2\lambda)},\\
c^{\rm O-Sp} &=& 1- \frac{3(4\lambda)^2}{\pi(\pi-2\lambda)}.
\end{eqnarray}
With the change of variables from $\lambda$ to $g$ related by (\ref{eqn:lg}),
we obtain the results (\ref{eqn:conformal}) and
(\ref{eqn:ccosp}). The mixed-boundary scaling index $t^{\rm O-Sp}$ defined by
\begin{equation}
c^{\rm O-Sp} = c^{\rm O-O} - 24 \; t^{\rm O-Sp}
\end{equation}
is therefore given by $t^{\rm O-Sp}=(3-2g)/4g=h_{1,2}$ in
the Kac formula (\ref{eqn:kac}). Using the notation $c_{\ell}$ for
the effective central charge for the eigenstate labelled
by $\ell$, we find from (\ref{eqn:on2}) and (\ref{eqn:alpha3}) that
\begin{eqnarray}
c_{\ell}^{\rm O-O} - c^{\rm O-O} &=& -24 \left[ \left(\smallfrac{1}{2}-
\smallfrac{2\lambda}{\pi}\right)\ell + \left(\smallfrac{1}{2}-\smallfrac{
\lambda}{\pi}\right)\ell^2 \right], \\
c_{\ell}^{\rm O-Sp} - c^{\rm O-Sp} &=& -24 \left[ -
\smallfrac{2\lambda}{\pi}\ell + \left(\smallfrac{1}{2}-\smallfrac{
\lambda}{\pi}\right)\ell^2 \right].
\end{eqnarray}
The scaling dimensions $X_{\ell}$ defined through the inverse correlation
lengths \cite{Cardy84a}
\begin{equation}
\xi_{\ell}^{-1}=
\log \frac{\Lambda_0}{\Lambda_{\ell}} \sim \frac{\pi \zeta X_{\ell}}{N}
\end{equation}
are then given by (\ref{eqn:xoo}) and (\ref{eqn:xosp}).

Let us now turn to the ground state for Sp-Sp boundaries, where there is
an imaginary Bethe ansatz root and the preceding analysis has to be slightly
modified. Firstly we noted numerically that the position of the imaginary
root reaches its asymptotic value much faster than $1/N$. We therefore
consider it to be fixed; and that this assumption does not affect the
finite-size corrections to the order in which we are interested. For
convenience we perform a shift $u_j\rightarrow u_{j-1}$ in the eigenvalue
expression (\ref{eqn:honeyev}) and associated Bethe ansatz equations. The
imaginary root is now $u_0 \approx \I\left(\smallfrac{\pi}{4}+\smallfrac{
\lambda}{2}\right)$ while the eigenvalue expression becomes
\begin{equation}
\Lambda_0^{\rm Sp-Sp} = \frac{\sinh(u_0+\smallfrac{3}{2}\I\lambda)
   \sinh(u_0-\smallfrac{3}{2}\I\lambda)}
   {\sinh(u_0+\smallfrac{1}{2}\I\lambda)\sinh(u_0-\smallfrac{1}{2}\I\lambda)}
\prod_{j=1}^{N-1}
\frac{\sinh(u_j+\smallfrac{3}{2}\I\lambda)\sinh(u_j-\smallfrac{3}{2}\I\lambda)}
{\sinh(u_j+\smallfrac{1}{2}\I\lambda)\sinh(u_j-\smallfrac{1}{2}\I\lambda)}.
\label{eqn:ssbae}
\end{equation}
The Bethe ansatz equations (\ref{eqn:logbae}) are now valid for $k=0,1,\ldots,
M=N-1$, with $I_k=k+1$.  The product term in (\ref{eqn:ssbae}),
which involves only real roots $u_j$ can now be treated as
before. The only difference is that the counting function $z_N(u)$ has to
be modified in order that $z_N(u_k)=I_k/N$ holds for the real roots $u_k$
in the shifted variables. We find that $\Phi(u)$ in (\ref{eqn:count}) has to
be replaced by
\begin{eqnarray}
\Phi_0^{\rm Sp-Sp}(u) & = & \phi_{\rm bound}(u,\lambda) -
  \phi(u+u_0,2\lambda)  + \phi(u+u_0,\lambda) \nonumber\\
  & + & \phi(2u,2\lambda)-\phi(2u,\lambda),
\end{eqnarray}
where unlike previously $u_0$ is no longer identified with $0$. With these
changes, the analysis of finite-size corrections then proceeds as before.
Since we have already obtained the surface energy from the excitations, we
restrict ourselves to the $O(N^{-2})$ terms. We find that Eq.\
(\ref{eqn:on2}) is unchanged. The sum rule (\ref{eqn:sumrule}) on the other
hand is now
\begin{equation}
\int_{\Lambda}^{\infty} \di u \rho_N(u) = \frac{1}{N}-\frac{2\lambda}{\pi N}
\end{equation}
due to $\Phi_0^{\rm Sp-Sp}(u)$. This results in the replacement of
(\ref{eqn:alpha3}) with
\begin{equation}
4 \pi^2 \alpha^2 = \left(1-\smallfrac{2\lambda}{\pi}\right)^{-1}
\left(\smallfrac{1}{2} - \smallfrac{2\lambda}{\pi}\right)^2.
\label{eqn:alpha4}
\end{equation}
Equations (\ref{eqn:on2}) and (\ref{eqn:alpha4}) then give rise to
\begin{equation}
c^{\rm Sp-Sp} = 1- \frac{3(\pi-4\lambda)^2}{\pi(\pi-2\lambda)},
\end{equation}
which is the same as for O-O boundaries. From Eqs.\ (\ref{eqn:on2}) and
(\ref{eqn:alpha3}) we can now calculate the effective central charge for
$\Lambda^{\rm Sp-Sp}_{\ell}$:
\begin{equation}
c_{\ell}^{\rm Sp-Sp} - c^{\rm Sp-Sp} = -24 \left[
\smallfrac{2\lambda}{\pi}\left(1-\smallfrac{2\lambda}{\pi}\right)^{-1}+
\left(-\smallfrac{1}{2}-\smallfrac{2\lambda}{\pi}\right)\ell +
\left(\smallfrac{1}{2}-\smallfrac{\lambda}{\pi}\right)\ell^2 \right].
\end{equation}
The scaling dimensions $X_{\ell}^{\rm Sp-Sp}$ are then given by
(\ref{eqn:xss}).

\section{Discussion}

In this paper we have studied the $O(n)$ loop model on the honeycomb lattice
with open boundaries from the point of view of the underlying Izergin-Korepin
$R$-matrix and its associated $K$-matrices. Three inequivalent sets of
integrable boundary weights and their corresponding Bethe ansatz solutions
were thus obtained. Two of these sets were of ``pure-type'' \cite{Yung95a}
and one
of mixed-type. One of the pure-type boundary weights (denoted here by O-O),
with the bulk and surface  couplings being the same, was previously
studied using the coordinate Bethe ansatz \cite{Batchelor93} and argued to
correspond to the $O(n)$ model at the ordinary surface transition. By
analysing the associated Bethe ansatz equations for the second set
of pure boundary weights (denoted here by Sp-Sp) we found agreement with
known results (some of which were conjectures) for the special surface
transition \cite{Burkhardt89,Fendley94}, which in the $n\rightarrow 0$
limit corresponds to the adsorption
transition for polymers. This justifies our claim to have found the critical
couplings for the $O(n)$ loop model on the honeycomb lattice at the
special transition and allows us to ``prove'' the abovementioned conjectures.
The set of mixed-type boundary weights therefore should correspond to the
$O(n)$ model at the mixed ordinary-special surface transition. Our Bethe
ansatz analysis gives a central charge and mixed-boundary scaling index
which are indeed in agreement with a recent result \cite{Burkhardt94} for
this transition, providing further evidence for our claim. In addition our
analysis has furnished the scaling dimensions for this transition.

It is interesting that the $K$-matrices for the $A_2^{(2)}$ model
have ``accounted for'' the ordinary
and special surface transitions in the manner described in this paper. This
leaves the interesting open problem of finding an integrable lattice $O(n)$
model which exhibits the extraordinary surface transition \cite{Binder83}
which corresponds to having fixed $O(n)$ spins at the boundary,
and for which conformal invariance predictions exist
\cite{Burkhardt89,Burkhardt94}. It is also an interesting
open problem to derive the physical boundary $S$-matrices, which constitute
a starting point for the techniques of \cite{Fendley94}, from
the $K$-matrices. This ought to be achievable using the Bethe ansatz analysis
performed here, along the lines of \cite{Grisaru94} for the XXX model, or
using the vertex operator approach \cite{Jimbo94}.

\vspace{10pt}
\noindent{\large\bf Acknowledgements}

\noindent
This work has been supported by the Australian Research Council.

\end{document}